\documentclass[11pt]{article}

\usepackage{graphicx}
\usepackage{cite}
\usepackage{amssymb}
\usepackage{amsmath}
\usepackage[colorlinks=true]{hyperref}
\def\beq{\begin{equation}}
\def\eeq{\end{equation}}
\def\ba{\begin{array}}
\def\ea{\end{array}}
\def\bea{\begin{eqnarray}}
\def\eea{\end{eqnarray}}

\def\sq2{\sqrt{2}}
\def\sqs{\sqrt{s}}
\def\ab1{{\rm ab}^{-1}}
\def\End{\end{document}}

\newcommand{\gm}{\gamma^\mu}
\newcommand{\smn}{\sigma^{\mu \nu}}

\newcommand{\gn}{\gamma^\nu}
\newcommand{\gnc}{\gamma_\nu}

\newcommand{\vphi}{\varphi}

\newcommand{\darrd}{\stackrel{\leftrightarrow}{D}}
\newcommand{\darrdk}{\stackrel{\leftrightarrow}{D^k}}

\begin{document}

\title{Top quark flavor changing couplings at a muon collider.}
\author{ Daniel Ak\'e\thanks{daniel.ake@cinvestav.mx}, \
  Antonio O. Bouzas\thanks{abouzas@cinvestav.mx} \
  and F. Larios\thanks{francisco.larios@cinvestav.mx,
  corresponding author.} 
\\
Departamento de F\'{\i}sica Aplicada,
CINVESTAV-M\'erida, \\ A.P. 73, 97310 M\'erida,
Yucat\'an, M\'exico}


\maketitle

\begin{abstract}
  There is growing interest in the development of a muon collider that
  would make it possible to produce lepton collisions at energies of
  several TeV.  Among others, there can be significant contributions
  to electroweak gauge boson, Higgs boson and top quark physics.  In
  this work we pay attention to the latter, in particular, effective
  flavor-changing (FC) top-quark interactions.  We discuss the flavor
  changing $t \bar q$ ($q=u,c$) production processes that can be a
  good probe of the dimension-six top quark four-fermion and
  fermion-boson operators in the SMEFT.  We consider all sixteen
  operators that can generate flavor-changing top quark couplings.
  After comparing with the current LHC bounds, we find potential limits
  three or four orders of magnitude stronger for four-fermion operators.
  Concerning fermion-boson couplings, for the tensor operators $Q_{uW}$
  and $Q_{uB}$ we obtain the highest sensitivity.  We also observe that
  the effective $W$ approximation (EWA) does not apply with $Q_{uB}$.
\end{abstract}


\section{Introduction}
Due to the unique combination of accesible high energy collisions
and a relatively clean environment, there has been renewed interest
in the future construction of a muon collider
(MuC) \cite{muoncoll,mucreport22,muonsmasher}.
One outstanding purpose of the MuC is that, in principle, it could
produce the Higgs boson at the resonance energy \cite{greco21}.
However, in recent years there is more interest in studying the
potential of a MuC at high energies to probe the Higgs boson
couplings \cite{mantani20,meade22,meade23,abbott22,yang20,chen22}.
In addition, there are also studies on the
top quark \cite{wang23,belyaev23},
the bottom quark \cite{greljo22, greljo23,profumo}, $W$ boson
scattering \cite{yang22}, and other processes
\cite{homiller21, homiller22, fridell23, yao23, Yin:2020afe}.

In this work, we discuss the potential sensitivities that the MuC
can reach at different energies for flavor changing (FC) dimension
six operators.  Besides the $\mu \mu tu$ contact terms, we consider
operators that generate effective couplings of the top quark and the
gauge bosons \cite{Larios:2006pb}. Our purpose is to make a
preliminary, but comprehensive assessment of all the possible FC top
quark couplings.
We find out which operators have little sensitivity and which ones
have the highest sensitivity.  Then, in future more in-depth studies
we can focus on the latter.  We also make a comparison with the
current limits obtained by the LHC measurements on FC top-quark
decays.

Concerning the current status of the MuC project, there is one energy,
of $10$ TeV that is considerably high and at the same time is seen
upon as realistic in terms of cost and technical feasibility
\cite{muoncoll}.  It should be able to achieve about
$10\; {\rm ab}^{-1}$ of integrated luminosity and would probably be
built after initial stages at lower energies are constructed.  In this
study, we are interested in this energy in particular, but we also
include other energies and their respective luminosities, as they are
presented in the recent literature.  In table~\ref{tab:lumin} we show
the four energies and their potential integrated luminosities that we
use.  They have been specifically proposed in
ref.~\cite{muonsmasher}. As seen from the table, the integrated
luminosity grows with the collision energy, and we expect this feature
to bring about stronger limits from top quark production.

\begin{table}[ht!]
  \centering
  \begin{tabular}{|c|c|c|c|c|}\hline
    ($\sqs$)(TeV)  & $3$ & $6$ & $10$ & $14$
  \\\hline
  ${\cal L} ({\rm ab}^{-1})$ & $1$ & $4$ & $10$ & $20$
\\\hline
  \end{tabular}                               
  \caption{Benchmark values for energy and luminosity of a
    high energy MuC \cite{muonsmasher}.}
  \label{tab:lumin}
\end{table}

The outline of the article is as follows.  In
section~\ref{sec:operators} we identify all the possible dimension-six
operators relevant for top-quark production at the MuC.  In
section~\ref{sec:production} we discuss the two most important modes
of production, at tree level and with the top quark as final state:
$\mu^+ \mu^- \to t \bar u$ and
$\mu^+ \mu^- \to t \bar u \nu_\mu \bar \nu_\mu$.  At this level of
approximation we make a simple estimate of the different
sensitivities, and identify the two operators with the highest
potential.  In section \ref{sec:hadr.chan} we make a more detailed
analysis of the second mode of production, focused on the operators
$Q_{uB}^{k3}$, $Q_{uB}^{3k}$, $Q_{uW}^{k3}$, $Q_{uW}^{3k}$, with $k=1,$ 2.
We consider the top hadronic decay channel and perform a signal
and background analysis to obtain limits on the corresponding
Wilson coefficients $C^{uB}_{k3,3k}$ and $C^{uW}_{k3,3k}$.
In section~\ref{sec:conclusions} we present our conclusions.
Finally, in appendix \ref{sec:app} we make some brief comments on the
definition of the dimension-six operator coefficients when going from
gauge- to mass-eigenstate fields. In addition, recent limits on
fermion-boson operators from the ATLAS collaboration are given in
appendix \ref{sec:lhc.bounds}.


\section{Operators with FC top quark couplings.}
\label{sec:operators}

We refer to the list of dimension-six effective operators
that is widely known as the Warsaw basis that first appeared in
ref.~\cite{grz10}.  This list was introduced in the basis of
gauge eigenstates (tables 2 and 3).  We want to work in the
mass eigenstate basis, so we actually refer to
ref.~\cite{feynrules} which is a follow-up of \cite{grz10}.
In \cite{feynrules} the redefinition from the gauge to the mass
eigenstate basis is thoroughly discussed and Feynman rules
in $R_\xi$-gauges are provided.
In table~2 there are nine fermion-boson operators that
involve up-type quarks, and in table~3 there are seven four-fermion
operators with up-type quarks and leptons.  
The operators in table~2 that generate top quark FC couplings with
gauge bosons are \cite{feynrules}:
\bea Q^{(pr)}_{u\vphi} &=& (\vphi^\dagger \vphi) (\bar {{q}_{p}'}
{{u}_{r}'} {\tilde \vphi}) \;\; \to \left[ Htu \right]
\; ,\nonumber \\
Q^{(pr)}_{\vphi u} &=& (\vphi^\dagger i {\darrd}_\mu \vphi) (\bar
{{u}_{p}'} \gm {{u}_{r}'}) \;\; \to \left[ Ztu \right]
\; ,\nonumber \\
Q^{(pr)}_{\vphi ud} &=& i({\tilde \vphi}^\dagger {\darrd}_\mu \vphi)
(\bar {{u}_{p}'} \gm {{d}_{r}'}) \;\; \to \left[ Wtd,\; Wub \right]
\; ,\nonumber \\
Q^{(1)(pr)}_{\vphi q} &=& (\vphi^\dagger i {\darrd}_\mu \vphi) (\bar
{{q}_{p}'} \gm {{q}_{r}'}) \;\; \to \left[ Ztu,\; Zub \right]
\; ,\label{eq:operators} \\
Q^{(3)(pr)}_{\vphi q} &=& (\vphi^\dagger i {\darrdk}_\mu \vphi) (\bar
{{q}_{p}'} \tau^k \gm {{q}_{r}'}) \;\; \to \left[ Ztu,\; Zub,\; Wtd,\;
  Wub \right]
\; ,\nonumber \\
Q^{(pr)}_{uG} &=& (\bar {{q}_{p}'} \smn T^a {{u}_{r}'}) {\tilde \vphi}
G^a_{\mu \nu} \;\; \to \left[ Gtu \right]
\; ,\nonumber \\
Q^{(pr)}_{uB} &=& (\bar {{q}_{p}'} \smn {{u}_{r}'}) {\tilde \vphi}
B_{\mu \nu} \;\; \to \left[ \gamma tu,\; Ztu \right]
\; ,\nonumber \\
Q^{(pr)}_{uW} &=& (\bar {{q}_{p}'} \smn {{u}_{r}'}) \tau^k {\tilde
  \vphi} W^k_{\mu \nu} \;\; \to \left[ \gamma tu,\; Ztu, \; Wtd,\; Wub
\right]
\; ,\nonumber \\
Q^{(pr)}_{dW} &=& (\bar {{q}_{p}'} \smn {{d}_{r}'}) \tau^k {\vphi}
W^k_{\mu \nu} \;\; \to \left[ \gamma ub,\; Zub,\; Wtd,\; Wub
\right],\nonumber \eea where, for clarity, we have indicated the
effective vertices generated by each operator in square brackets.
We are interested in the case where
one of the flavor indices $pr$ is equal to 3 and the other one is 1 or
2.  Due to the high energy of the processes of interest the charm
quark can be taken as massless and there is no difference between
$t\bar u$ and $t\bar c$ production.  Therefore, we will usually refer
to specific flavor indices $13$ and $31$, with the understanding that
our results are equally valid for $23$ and $32$.

The operators in table~3 that generate contact terms of up quarks
and leptons are\cite{feynrules}:
\bea
Q_{lq}^{(1)} &=& (\bar {{\ell}_{p}'} \gnc {{\ell}_{r}'})  \;
(\bar {{q}_{s}'} \gn {{q}_{t}'}) \; ,
\nonumber \\
Q_{lq}^{(3)} &=&
(\bar {{\ell}_{p}'} \gnc \tau^k {{\ell}_{r}'})  \;
(\bar {{q}_{s}'} \gn \tau^k {{q}_{t}'}) \; ,
\nonumber \\
Q_{eu}^{} &=&
(\bar {{e}_{p}'} \gnc {{e}_{r}'})  \;
(\bar {{u}_{s}'} \gn {{u}_{t}'}) \; ,
\nonumber \\
Q_{lu}^{} &=&
(\bar {{\ell}_{p}'} \gnc {{\ell}_{r}'})  \;
(\bar {{u}_{s}'} \gn {{u}_{t}'}) \; ,
\label{eq:operators4f} \\
Q_{qe}^{} &=&
(\bar {{e}_{p}'} \gnc {{e}_{r}'})  \;
(\bar {{q}_{s}'} \gn {{q}_{t}'}) \; ,
\nonumber \\
Q_{lequ}^{(1)} &=&
(\bar {{\ell}_{p}^{'j}} {{e}_{r}'})
\; \epsilon_{jk} \;
(\bar {{q}_{s}^{'k}} {{u}_{t}'}) \; ,
\nonumber \\
Q_{lequ}^{(3)} &=&
(\bar {{\ell}_{p}^{'j}} \sigma^{\alpha \beta} {{e}_{r}'}) 
\; \epsilon_{jk} \;
(\bar {{q}_{s}^{'k}} \sigma_{\alpha \beta} {{u}_{t}'}) \; .
\nonumber
\eea
Where, in operator $Q_{qe}$ we have changed the order of the quark
and lepton terms so as to keep the same arrangement of all the
other operators.  As in the case of fermion-boson operators,
we will be interested in the flavor indices $prst=2213$ that are
related to $\mu^+ \mu^- \to \bar u t$.  The numerical results for
$\bar t u$, $\bar c t$ and $\bar t c$ production are the same.

Looking at equation (\ref{eq:operators}),
let us observe that two operators generate FCNC couplings
with up type quarks as well as with down type quarks:
$Q_{\vphi q (1)}$ and $Q_{\vphi q (3)}$.  As far as FC charged
currents are concerned, operators $Q_{\vphi ud}$, $Q_{\vphi q (3)}$,
$Q_{uW}$ and $Q_{dW}$ also generate couplings with or without the top
quark, depending on what combination of indices is considered: either
$13$ or $31$.
In one case the operator coefficient can only be probed via top quark
production or top decay, and in the other case we could have bottom
quark production or B meson decay as another good probe for the
coefficient.  In particular, concerning operators $Q^{(1)}_{\vphi q}$ and
$Q^{(3)}_{\vphi q}$, the following combinations are usually defined in
the literature \cite{durieux}: \bea
Q_{\vphi q(-)} &\equiv& Q^{(1)}_{\vphi q} - Q^{(3)}_{\vphi q} \; , \nonumber \\
Q_{\vphi q(+)} &\equiv& Q^{(1)}_{\vphi q} + Q^{(3)}_{\vphi q} \; .
\eea Where the $Q_{\vphi q(-)}$ operator generates $Ztu$ and
$Wtd,\,Wub$ couplings, whereas $Q_{\vphi q(+)}$ generates $Zbd$ and
$Wtd,\,Wub$.  In previous studies, it is assumed that
$Q_{\vphi q(+)}$ will be strongly bound by some down-quark FCNC
process and therefore not included in a discussion of top quark
production or decay \cite{durieux}.  While this is clearly justified,
we point out two things: 1) in a future global analysis it may still
be important to consider the contribution from $Q_{\vphi q(+)}$ if
the sensitivity is good enough for top production, and 2) as a matter
of fact, the other operator $Q_{\vphi q(-)}$ also generates a $Wub$
coupling, which contributes to B meson decay at tree level so it also
gets strong bounds \cite{greljo23}.
Concerning four-fermion operators, we will use a similar combination
\cite{durieux}:
\bea
Q_{lq-} = \frac{1}{2} (Q_{lq}^{(1)} - Q_{lq}^{(3)}) \; ,\nonumber
\eea
that separates the vertices $\mu \mu ut$ and $\mu \mu db$.

The Lagrangian in terms of interaction eigenstates is
written as (omitting flavor indices):
\bea
{\cal L} &=& {\cal L}_{\rm SM}
\; +\; \frac{1}{\Lambda^2}{C'}_{X} Q_{X {\rm Hermitian}}
\nonumber \\
&+& \frac{1}{\Lambda^2} \left( {C'}_{Y} Q_{Y} \; +
\; {C'}^{\dagger}_Y  Q^{\dagger}_{Y} \right) \; ,
\label{eq:smeft}
\eea
where we will set the mass scale $\Lambda =1$ TeV for convenience.
It is worthwhile to bear in mind that in this effective Lagrangian
framework, the amplitudes we compute are valid as long as the
process takes place at energies (well) below the scale $\Lambda$.
For instance, at $\sqs = 3$ TeV $\Lambda$ should be about $10$ TeV
or higher.  Of course, we do not know the scale of new physics and
in presenting limits it would be appropriate to refer to
$C_Q/\Lambda^2$ instead of just the coefficient $C_Q$.  However,
it is also customary to set $\Lambda =1$TeV and the reader can
infer that the number presented as a limit would be given in
units of ${\rm TeV}^{-2}$ for $C_Q/\Lambda^2$
(see ref.~\cite{durieux} for example).
The Lagrangian of eq.~(\ref{eq:smeft}) is the basis of what is
known as the Standard Model Effective Field Theory (SMEFT).
For a recent general and in-depth discussion of this topic
we would like to refer to the article~\cite{isidori23}
by G.~Isidori, F.~Wilsch and D.~Wyler, where issues about the
validity of this framework are reviewed.
For instance, if we were
to compute the contribution of one of these top quark operators to
the amplitude of the $b\to s\gamma$ decay, we would refer to
a matching scale $\Lambda =M_W$ equal to the $W$ boson mass
(see fig.~10 in \cite{isidori23}).
The effective operator of the Light Effective Field Theory
(LEFT) that is used to compute the decay width receives contributions
from operators of the SMEFT at the matching scale.  Then, a
renormalization group running is performed to obtain the value
of this operator at the $m_b$ scale, which is the relevant scale for
this process.   Similarly, there is a Beyond Standard Model theory
that supersedes the SMEFT at a certain matching scale $\Lambda_{\rm BSM}$.
Of course, we do not know what is $\Lambda_{\rm BSM}$.  When a
phenomenological study is done we have to refer to a certain
energy scale that is appropiate for the study, one that is
supposed to be below $\Lambda_{\rm BSM}$.
The limits that we will obtain for these operators are assumed to
be taken at the energy scale of the production process, which
for the MuC we have chosen to be one of the four values in particular:
$3,6,10$ and $14$ TeV.  In each case we are assuming that
$\Lambda_{\rm BSM}$ is above that energy in particular.
Strictly speaking, when we compare our limits with the current
limits from LHC measurements we should be aware that there will be
a renormalization group running from the scale of the collision
energy down to the appropiate energy scale of the LHC.
This may result in a variation of the order of $10\%$.  We are
not interested in making such detailed analysis in this study.

Operators $Q_{\vphi u}$, $Q_{\vphi q(-)}$ and $Q_{\vphi q(+)}$ are Hermitian.
The next step in our analysis is to re-write the operators and their
Wilson coefficients in the mass eigenstate basis. For this we refer to
ref.~\cite{feynrules} where all the dimension-six operators are
considered and the transition from interaction to mass eigenstates is
discussed. We follow the treatment in \cite{feynrules}, with some
minor variations discussed below in appendix \ref{sec:app}.  It is
useful to establish the number of free parameters in our list of 9
fermion-boson operators. Let us consider the flavor index $a=1$.
We have three Hermitian operators and their coefficients must satisfy
the condition $C_{13}=C^*_{31}$.  Otherwise, $C_{13}$ and $C_{31}$ are
independent.  We thus have $3+12=15$ complex free parameters for $a=1$.
For simplicity, however, in this study we consider them to be real
numbers.
Therefore, we are not considering CP violation effects in this study.
In principle, all operators can give a contribution to some single
top production process at the MuC.  In trying to be comprehensive,
we have checked the potential for each and every coefficient.
It turns out that, only nine independent real couplings ($3+6$) will
actually be relevant to our analysis.   The reason is that for three
operators, $Q^{\vphi ud}$, $Q^{uG}$ and $Q^{dW}$ the contribution to any
$t\bar u$ production process will be either zero (by taking $m_d=0$)
or negligible.  So that their potential MuC limits will be two or
more orders of magnitude weaker than current limits, and they
will not appear in any of the tables where we show our results.


Let us now refer to four-fermion operators.  They give rise to the
two-to-two process $\mu^+ \mu^- \to t{\bar u}$ (or ${u\bar t}$).
(The lepton flavor indices are always $pr=22$ and there is no
need to include them.)
There are seven operators in equation (\ref{eq:operators4f}) with
the first five being Hermitian.  So we have $4+4$ independent
coefficients: $C^{lq-}_{1+3}$, $C^{eu}_{1+3}$, $C^{lu}_{1+3}$,
$C^{qe}_{1+3}$, $C^{lequ1}_{13}$, $C^{lequ1}_{31}$, $C^{lequ3}_{13}$ and
$C^{lequ3}_{31}$ \cite{durieux}.

Given the energies of order TeV consider here, we have that
terms or order $m_t/\sqs$ are small.  It is reasonable
to assume massless fermions in the calculation of the
cross section.  In this case the contact terms give rise to
very simple helicity amplitudes and associated production
cross sections.  By using the expressions provided in a previous
work \cite{emu22} we obtain:

\bea
\frac{\sigma_{\mu^+ \mu^- \to \bar u t}}{\sigma_{1234}} &=&
|C_{1+3}^{lq-}|^2 + |C_{1+3}^{eu}|^2 +
|C_{1+3}^{lu}|^2 + |C_{1+3}^{qe}|^2 \nonumber \\
&+& \frac{3}{4} ( |C_{13}^{lequ1}|^2 + |C_{31}^{lequ1}|^2 ) +
4 ( |C_{13}^{lequ3}|^2 + |C_{31}^{lequ3}|^2 ) \;.
\label{eq:cross4f}
\eea

Where $\sigma_{1234} = s/(16\pi \Lambda^4)$ is the total
cross section coming from the helicity structure
$|[12][34]|^2 =s^2$ discussed in Ref.~\cite{emu22}.
Numerically, for $\Lambda =1$TeV and $\sqs =3$TeV, we get
a rather high $\sigma_{1234} \simeq 70$pb.
We have computed $\sigma_{1234}$ with $m_t$ not zero and 
the numerical difference is less than $3\%$.  We will show
limits for the four-fermion coefficients in the next section.


\section{Single top production at the MuC.}
\label{sec:production}

\begin{figure}[ht!]
\centering
\includegraphics[scale=0.4]{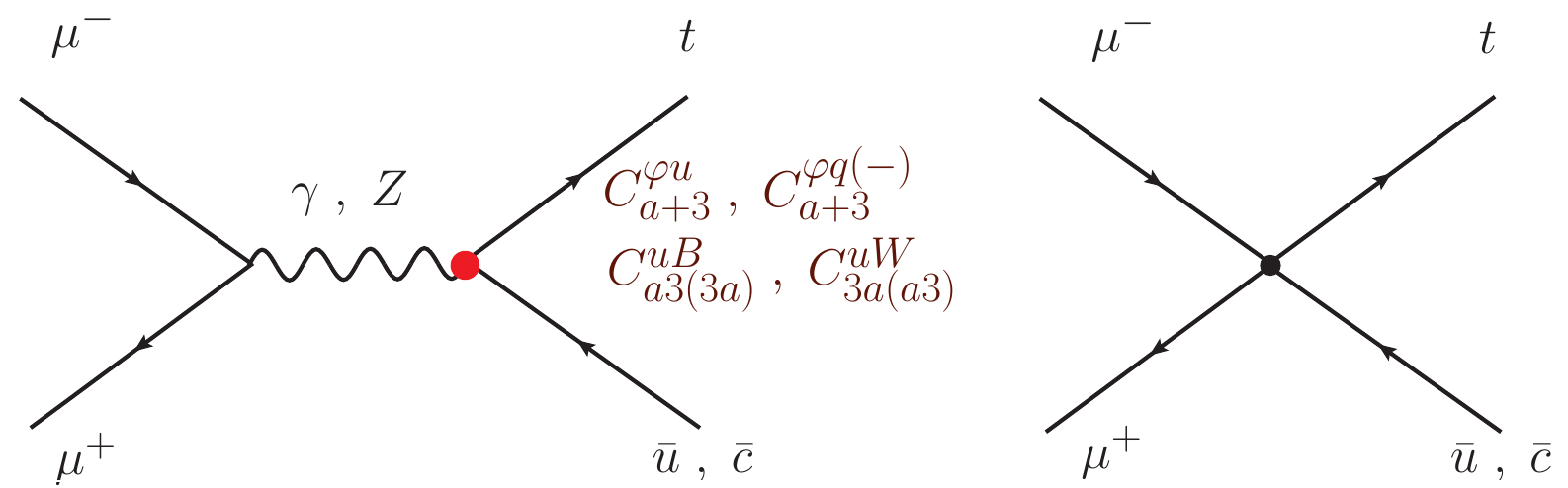}  
\caption{The FC process $\mu^- \mu^+ \to t\bar u \; ({\bar c})$
  as generated by the $tq\gamma \; (Z)$ vertex or
  a $\mu \mu t q$ contact term.}
  \label{fig:mmtu}
\end{figure}

In the SM, production of single top at a lepton collider comes from
the process $\ell^+ \ell^- \to t{\bar b}W^- + {\bar t}bW^+$ and it can
be used to probe the diagonal terms of the Wilson coefficients
\cite{escamilla}.  Single-top production that involves FC couplings
can be given by $\mu^+ \mu^- \to t{\bar u} + u{\bar t}$ as shown in
fig.~\ref{fig:mmtu}.  This top production mode is possible even with a
low collision energy of $240$ GeV (as in the CEPC), and good
sensitivity can in principle be achieved with very high integrated
luminosity \cite{zhang19,gori23,iranies14,iranies21}.  The simplest
diagram is given by the $\mu \mu t q$ contact term. For the LHC,
one recent study be found in ref.~\cite{wudka21}.
For the MuC the authors in ref.~\cite{sun2023} consider $Q^{(1)}_{lq}$
for the $\mu \mu tc$ coupling.
Other recent studies for FC top couplings at the HL-LHC can be found
in \cite{cremer23,nomura22}.
We will also consider associated production modes of $t\bar u \gamma$
and $t\bar u Z$ as well as 
$\mu^+ \mu^- \to \nu_\mu \bar {\nu_\mu} t{\bar u}$ where only
fermion-boson operators have a potential for good sensitivity.

In order to make an estimate of the potential MuC individual limits
for each operator coefficient, let us use a simple criterion: 
we find the value of the coefficient at which there are a total of
one thousand $t\bar u + u\bar t$ (that is $500+500$) events.
Given that there is a different expected luminosity at each energy
(see table \ref{tab:lumin}) there are four minimal cross sections
depending on the energy.  For instance, at $\sqs =3$ TeV we would
have $\sigma_{\rm min} = 500$ ab.

\subsection{Limits on four-fermion operators from
  $\mu^+ \mu^- \to t{\bar u}$.}

From the cross section result in equation (\ref{eq:cross4f}) we
can obtain the potential limits of the four-fermion coefficients.
They are shown in table~\ref{tab:4flimits} below.

\begin{table}[ht!]
  \centering
  \begin{tabular}{|c|c|c|c|c|}\hline
    ${\;\;}$  & $3$TeV & $6$TeV & $10$TeV & $14$TeV
  \\\hline
  $10^4 \times |C^{lq-}_{1+3}|\leq$ &
  $26.8$ & $6.7$ & $2.5$ & $1.3$ 
  \\\hline
   $10^4 \times |C^{lequ1}_{13}|\leq$ &
  $30.9$ & $7.7$ & $2.9$ & $1.5$
  \\\hline
  $10^4 \times |C^{lequ3}_{13}|\leq$ &
  $13.4$ & $3.4$ & $1.3$ & $0.64$
\\\hline
  \end{tabular}                               
  \caption{Limits on the 4-fermion coefficients from
    $\mu^+ \mu^- \to t\bar u$.}
  \label{tab:4flimits}
\end{table}

Notice that we have only included three coefficients in
table~\ref{tab:4flimits} as the limits of the other five
coefficients are the same as seen from eq.~(\ref{eq:cross4f}).
These are very stringent limits of order $10^{-4}$
(for $\Lambda = 1$TeV).  They are three to four orders of magnitude
smaller than limits from LHC and LEP data, which are of order
1\cite{durieux,atlas2023}.

\subsection{Limits on fermion-boson operators from
    $\mu^+ \mu^- \to t{\bar u}$.}

For the case of fermion-boson operators we will consider three
production modes:
first, $\mu^+ \mu^- \to t{\bar u}$ that involves FCNC couplings only;
second, the associated production with an additional $\gamma$ or $Z$;
and third, $\mu^+ \mu^- \to t{\bar u} \nu_\mu \bar \nu_\mu$ that involves
diagrams with $WW$ fusion where FC charged-current couplings may play
an important role.  We compute the contribution from each operator
coefficient, obtain limits on the couplings and compare with the
current bounds from the LHC.  We compute the tree-level cross section
for FC top production at the MuC with the matrix-element Monte Carlo
generator MadGraph5\_aMC@NLO (henceforth MG5) version 3.5.1
\cite{alw14}.  We implement the dimension-six operators
(\ref{eq:operators}) in MG5 by means of FeynRules 2.3.49
\cite{all14}. Furthermore, we analyze the events generated by MG5 with
ROOT 6.26 \cite{root}. We also used the program CalcHEP for
crosschecks of our results \cite{calchep}. For the numerical results
in this work we use: ${\alpha}^{-1}=127.9$,
${{\sin}^2{\theta_W}}=0.2337$, $M_Z=91.187{\rm GeV}$,
$m_t=172.5{\rm GeV}$, $m_b=4.5{\rm GeV}$, $m_u=m_d = m_s = m_c =0$ and
$\Lambda = 1{\rm TeV}$.

We will impose the following basic cuts on the transverse momentum and
the pseudo-rapidity of the light jet in all the production processes
in this section,
\bea 
p_T (j) \geq 10 {\rm GeV} \; , \;\;\; |y(j)| \leq 3 \; .
\label{eq:cutbasic}
\eea
The cross sections for $\mu^+ \mu^- \to t{\bar u}$, at the energies
$(3,6,10,14)$ TeV considered here, and as produced by the
operators $Q_{\vphi u}$ and $Q_{\vphi q(-)}$, are actually very low
(of order ab):
\bea
\sigma (t\bar u) /|C^{\vphi q(-)}_{1+3}|^2 &=&
120 \; ,\; 30 \; ,\; 11 \; ,\; 6.0 \;\; {\rm ab},
\nonumber \\
\sigma (t\bar u) /|C^{\vphi u}_{1+3}|^2 &=&
30 \; ,\; 7.4 \; ,\; 2.7 \; ,\; 1.4 \;\; {\rm ab}.
\label{eq:tulow1}
\eea

From the numbers shown in eq.~(\ref{eq:tulow1}) we can already see
that the potential limits for operators $Q_{\vphi q(-)}$ and $Q_{\vphi u}$
must be higher than 1.  Indeed, based on the minimal cross sections
required for $1000$ ($t\bar u + \bar t u$) events
we get $C^{\vphi q(-)}_{1+3}< 2$ and
$C^{\vphi u}_{1+3} < 4$ at all energies.  This sensitivity is about
one order of magnitude lower than the current limits reported by
ATLAS: $|C_{1+3}^{\vphi q(-)}|$ and $|C_{1+3}^{\vphi u}|$ get limits from
$t\to uZ$ of $0.1$ and $0.2$ respectively \cite{atlas2023} (see
appendix \ref{sec:lhc.bounds}).

In contrast, the cross section $\sigma (\mu^+ \mu^- \to t{\bar u})$
is large enough for the tensor operators $Q_{uB}$ and $Q_{uW}$.
It is constant in energy, and of order fb:
\bea
\sigma (t\bar u) /|C^{uB}_{13(31)}|^2 &=& 75.3 \;{\rm fb}.
\nonumber \\
\sigma (t\bar u)/|C^{uW}_{13(31)}|^2 &=& 49.4 \;{\rm fb}.
\label{eq:tuhi}
\eea
The reason for such behavior in energy is in the tensor coupling
term $\sigma^{\mu \nu} p_\nu$.  We then will have $Q_{uB}$ and $Q_{uW}$
as the two fermion-boson operators with the highest sensitivity
for  $\mu^+ \mu^- \to t{\bar u}$.

In table~\ref{tab:hilimits} the limits on the coefficients are shown,
as obtained from the latest LHC bounds on $Br(t\to Vq)$
\cite{atlas2023} (see appendix \ref{sec:lhc.bounds}).
We will use the (strongest) bounds from $t\to u\gamma$ to compare with
the potential limits at the MuC.
\begin{table}[ht!]
  \centering
  \begin{tabular}{|c|c|c|c|c|}\hline
   LHC  & $|C^{uB}_{13(31)}|\leq$ & $|C^{uB}_{23(32)}|\leq$ &
   $|C^{uW}_{13(31)}|\leq$ & $|C^{uW}_{23(32)}|\leq$
  \\\hline
$t\to u\gamma , c\gamma$ & $0.04$ & $0.08$ & $0.07$ & $0.15$ 
   \\\hline
 $t\to uZ , cZ$ & $0.23$ & $0.34$ & $0.13$ & $0.19$ 
\\\hline
  \end{tabular}                               
  \caption{Limits on $|C^{uB}|$ and $|C^{uW}|$
 from LHC top decay measurements \cite{atlas2023}.   }
  \label{tab:hilimits}
\end{table}
In table~\ref{tab:tulimits} we show the limits we obtain from
the minimal cross sections required for $500$ $t\bar u$ events.
We are showing the ratios
$R=|C^{\rm max}_{\mu {\rm MuC}}| / |C^{\rm max}_{\rm LHC}|$
to establish the comparison.  At the higher energies $10,14$ TeV
the constraints are stronger by a factor of 2 or 3.

\begin{table}[ht!]
  \centering
  \begin{tabular}{|c|c|c|c|c|}\hline
    ${\;\;}$  & $3$TeV & $6$TeV & $10$TeV & $14$TeV
  \\\hline
  $10^2 \times |C^{uB}_{13(31)}|\leq$ &
  $8.9{\;}(R=2.1)$ & $4.4{\;}(R=1.0)$ &
  $2.0{\;}(R=0.65)$ & $1.4{\;}(R=0.45)$ 
  \\\hline
  $10^2 \times |C^{uW}_{13(31)}|\leq$ &
  $11{\;}(R=1.4)$ & $5.5{\;}(R=0.72)$ &
  $2.5{\;}(R=0.46)$ & $1.7{\;}(R=0.32)$
\\\hline
  \end{tabular}                               
  \caption{Limits on the coefficients from $\mu^- \mu^+ \to t\bar u$.
    The ratios $R=|C^{\rm max}_{\mu {\rm MuC}}| / |C^{\rm max}_{\rm LHC}|$
    are shown.}
  \label{tab:tulimits}
\end{table}

Let us notice that in obtaining the results of equations (\ref{eq:tulow1})
and (\ref{eq:tuhi}) we have ignored contributions from the SM.  Because of
the GIM mechanism \cite{gim} there are no FCNC couplings in the SM at tree
level.  At the one loop level the mechanism does not quite hold because of
the difference in up type quark masses.  If the masses $m_u$, $m_c$ and $m_t$
were equal there would be no FCNC couplings either (see for instance
\cite{altmannshofer}).  The decay $b\to s \gamma$ (or $b\to d\gamma$) is
sizeable enough to be observable because of the size of the top quark mass.
However, the decay $t\to u \gamma$ (or $t\to c\gamma$) is very small, of
order $10^{-13}$ because $m_b$ is large compared to $m_s$ and $m_d$ but
still much smaller than $m_t$.  To give us an idea of how large is the
SM contribution to the Wilson coefficients let us take a look at the
branching ratios \cite{Larios:2006pb}:
\bea
Br(t\to u V) &=& f_Q \times 10^{-3} |C^Q_{13,31}|^2 \; . \nonumber
\eea
With $f_Q \simeq (0.5 \; ,\; 11)$ for $V=(H,G)$ and
$Q=(Q_{u\vphi} ,Q_{uG})$;
$f_Q \simeq (2.0 \; ,\; 6.4)$ for $V=\gamma$ and
$Q=Q_{uW} ,Q_{uB}$; 
$f_Q \simeq (6.7 \; ,\; 1.7 \; ,\; 3.8 \; ,\; 1.2)$ for $V=Z$ and
$Q=Q_{\vphi q(-)}, Q_{\vphi u}, Q_{uW} , Q_{uB}$.
So we see that $B_{\rm SM} (t\to u\gamma) \sim 10^{-13}$ means that the
SM contribution to the coefficients is of order $10^{-5}$.  This
number is at least three orders of magnitude smaller than the
potential MuC limits we are obtaining.  The SM contribution is
negligible.


\subsection{Associated production $\mu^- \mu^+ \to t{\bar u} V$,
  with $V=\gamma$, $Z$, $g$.}
\label{sec:tua.tuz}

\begin{figure}[ht!]
\centering
\includegraphics[scale=0.4]{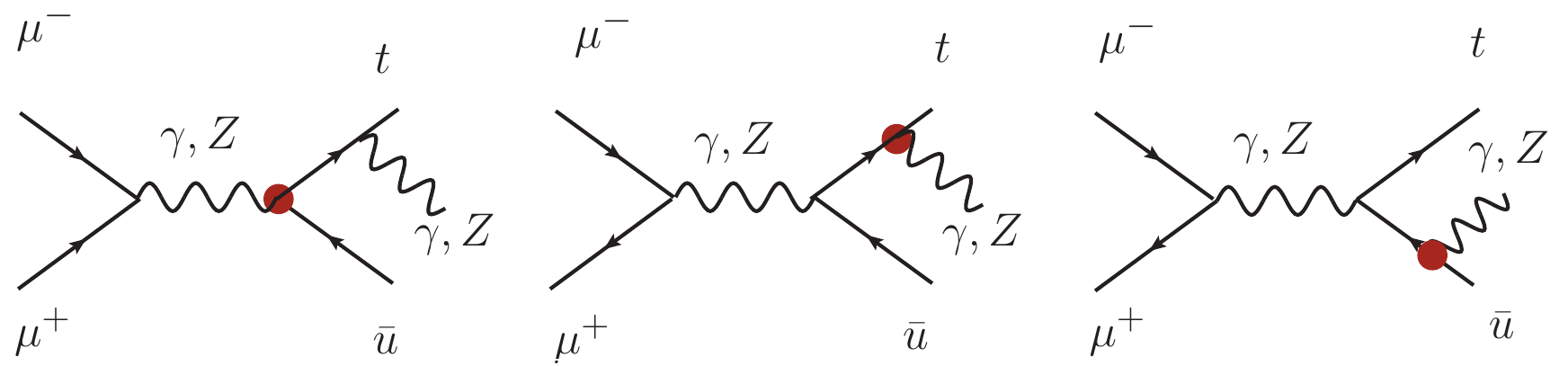}  
\caption{The FC process $\mu^- \mu^+ \to t\bar u \; +\gamma (Z)$.
  The FC coupling involves either the internal or the external
  boson line.}
  \label{fig:mmtuz}
\end{figure}

Let us now consider the associated production modes $t{\bar u} \gamma$
and $t{\bar u} Z$ that involve $\mu^+ \mu^-$ annihilation.
In fig.~\ref{fig:mmtuz} we show three sample diagrams.
Notice that the FC coupling can appear in the internal
line as in $t\bar u$ production and then the $\gamma$ or $Z$
boson is emitted via the regular SM coupling.  There is another
source of flavor change when the photon (or $Z$) is emitted
via the dimension six operator.  For the tensor operators the
coupling is proportional to the energy of the $\gamma$ or $Z$ and we
expect some effect of increasing amplitude values as the energy of the
collider is also increased.

For $t\bar u \gamma$ we impose,
in addition to eq.~(\ref{eq:cutbasic}), the following cuts:
\bea
p_T (\gamma) \geq 10 {\rm GeV} \; , \; |y(\gamma)| \leq 3 \; ,\;
\Delta R (\gamma ,j) \geq 0.4 \; .
\label{eq:cutgamma}
\eea

As in the case of $t{\bar u}$ production, the cross section numbers
produced by the operators $Q_{\vphi u}$ and $Q_{\vphi q(-)}$ are rather
low:
\bea
\sigma (t\bar u \gamma [t\bar u Z])/|C^{\vphi q(-)}_{1+3}|^2 &=&
20\;[160] ,\; 6.6\; [44] ,\; 2.8\; [17] ,\;
1.6\; [7.0] \; {\rm ab},
\nonumber \\
\sigma (t\bar u \gamma [t\bar u Z])/|C^{\vphi u}_{1+3}|^2 &=&
5.8\; [36] ,\; 1.8\; [9.5] ,\; 0.74\; [3.6] ,\;
0.42\; [1.9]\; {\rm ab}.
\label{eq:tulow2}
\eea
Again, the coefficients would have to take on values of about $10$
in order to produce enough events in $t\bar u \gamma$ and of about
4 or 5 in $t\bar u Z$ production.  We conclude that, as before, the
MuC has a low sensitivity for $Q_{\vphi u}$ and $Q_{\vphi q(-)}$ in this
case.


For operators $Q_{uB}$ and $Q_{uW}$ the cross sections produced are:
\bea
\sigma (t\bar u \gamma [t\bar u Z])/|C^{uB}_{13(31)}|^2 &=&
10\; [18] ,\; 12\; [70] ,\; 14\; [197] ,\;
15\; [385]\; {\rm fb},
\nonumber \\
\sigma (t\bar u \gamma [t\bar u Z])/|C^{uW}_{13(31)}|^2 &=&
6.5 \; [12] ,\; 8.2\; [46],\; 9.3\; [129] ,\;
10 \; [252]\; {\rm fb}.
\label{eq:tugammaz}
\eea
To set limits we now require the minimal cross sections to be
twice the values requested before, that is $2000$ events
($t\bar u + \bar t u$).  The reason is
that we expect the background to significantly increase when
considering the emission of a photon, and even more so when
considering the two additional fermions from $Z$ decay.

In table~\ref{tab:tuzlimits} we show the limits associated with these
production modes.  If we compare with the limits from $t\bar u$ in
table~\ref{tab:tulimits}, we can see that the limits are now weaker
coming from $t\bar u \gamma$ and are of similar size when coming
from $t\bar u Z$.  Of course, these are rough estimates but we
believe that $t\bar u \gamma$ and $t\bar u Z$ production could
be useful in a global analysis because four-fermion operators
should yield a lower cross section in this mode.
The emission of $\gamma , Z$ with four-fermion couplings must come
from regular SM couplings.

Another operator with low sensitivity is $Q_{uG}$.  The only process
at the MuC at tree level to be considered is the radiative associated
production of single top: $\mu^+ \mu^- \to t {\bar u}g$.  We have
computed the cross sections at the energies of table~\ref{tab:lumin},
with $C_{13(31)}^{uG}=1$ and applying some basic cuts to avoid soft
and collinear regions, as in $t {\bar u} \gamma$, with the result of
a constant $\sigma (\mu^+ \mu^- \to t {\bar u}g) = 62$ ab for the
entire range of energies $\sqs = 3-14$TeV.  The energy dependence of
the tensor coupling is proportional to $\sqs$ and that is the reason
why the cross section remains essentially constant with increasing
energy. This small value for the cross section suggests a
sensitivity to $C_{13(31)}^{uG}$ somewhat higher than of order 1.
This is much lower than the current limits obtained at the LHC.
Indeed, $Q_{uG}$ is very well probed with the
single top quark production $gu (gc) \to t$ as well as the decay
process $t\to ug (cg)$ at the LHC.  The current limits from non
observation of $t\to ug (cg)$ decay as reported by ATLAS
are $|C_{13(31)}^{uG}| \leq 0.074$ and $|C_{23(32)}^{uG}| \leq 0.18$
\cite{atlas2023} (see appendix \ref{sec:lhc.bounds}).
We remark, however, that $Q_{uG}$ may still play an important role
in a lepton collider study in the framework of a global analysis
involving QCD NLO corrections \cite{durieux}.

\begin{table}[ht!]
  \centering
  \begin{tabular}{|cc|c|c|c|c|}\hline
    ${\;}$ & ${\;}$  & $3$TeV & $6$TeV & $10$TeV & $14$TeV
  \\\hline
$t{\bar u} \gamma$: & $10^2 \times |C^{uB}_{13(31)}|\leq$ &
  $32{\;}(R=8.0)$ & $15{\;}(R=3.8)$ & $8.4{\;}(R=2.1)$
  & $5.8{\;}(R=1.5)$ 
  \\
  $t{\bar u}Z$: & ${\;\;}$ &  $24{\;}(R=6.0)$ &
  $6.0{\;}(R=1.5)$ & $2.3{\;}(R=0.58)$ & $1.2{\;}(R=0.3)$
  \\\hline
$t{\bar u} \gamma$: & $10^2 \times |C^{uW}_{13(31)}|\leq$ &
  $39{\;}(R=5.6)$ & $18{\;}(R=2.6)$ & $10{\;}(R=1.4)$ &
  $7.1{\;}(R=1.0)$
  \\
  $t{\bar u}Z$: & ${\;\;}$ & $29{\;}(R=4.1)$ &
  $7.4{\;}(R=1.1)$ & $2.8{\;}(R=0.4)$ & $1.4{\;}(R=0.2)$
  \\\hline
  \end{tabular}                               
  \caption{Limits on the coefficients from
$\mu^- \mu^+ \to t\bar u V$ with $V=\gamma ,Z$ obtained with the
    minimal cross sections required for $1000$ $t\bar u$ events.}
  \label{tab:tuzlimits}
\end{table}

\subsection{The process
$\mu^- \mu^+ \to t{\bar u} \nu_\mu \bar \nu_\mu$.}
\label{sec:mmtunn}

\begin{figure}[ht!]
\centering
\includegraphics[scale=0.4]{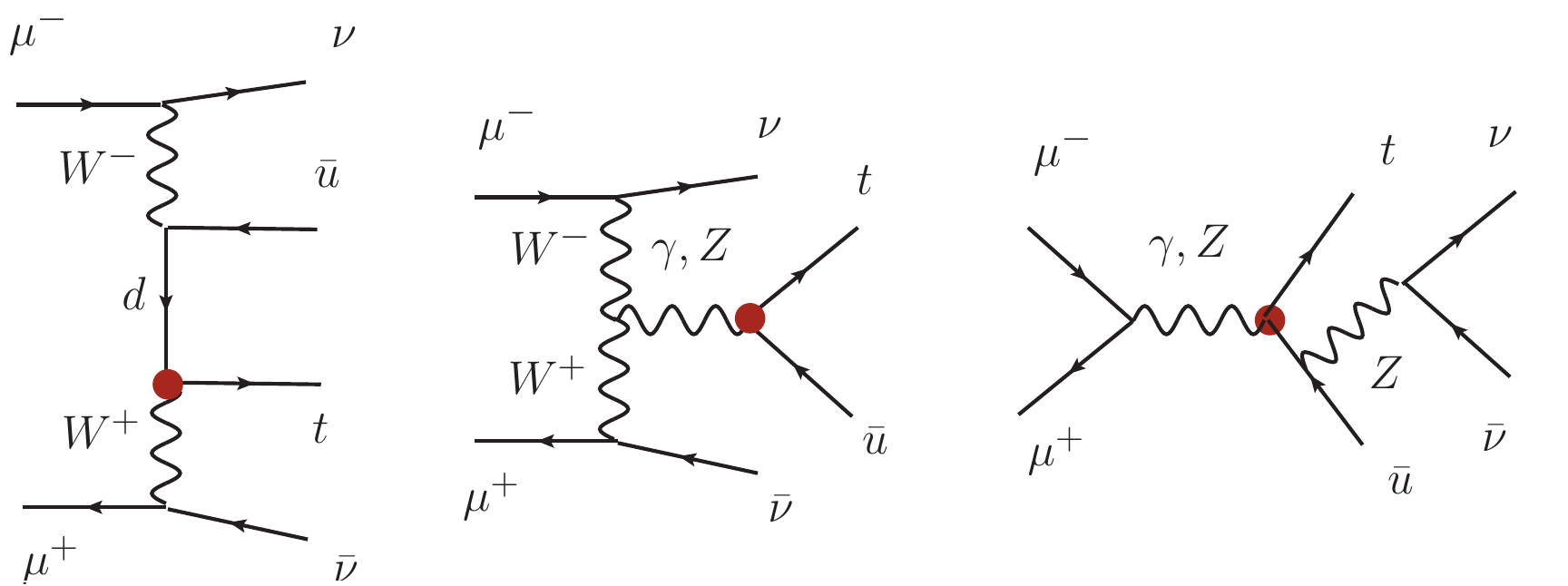}  
\caption{The FC process $\mu^- \mu^+ \to t\bar u \; \nu \bar \nu$:
  three sample diagrams.}
  \label{fig:mmtunn}
\end{figure}

This process offers a richer phenomenology than
the previous $t{\bar u}$ or $t{\bar u}\gamma (Z)$.
In the SM, one can have the three t-channel $WW$ fusion diagrams like
the one in the left hand side of fig.~\ref{fig:mmtunn}.
When we add them together, they cancel each other.
This is due to the unitarity of the CKM matrix
--the GIM mechanism.  Therefore, the SM contribution is zero at tree
level. When we include the dimension six FC couplings
like $Wtd$ or $Wub$ there is no more cancellation and we can get a
non-zero cross section.   At high energies the contribution from these
$WW$ fusion diagrams becomes significant or even dominant, and then a
computational simplification can be used, known as the effective $W$
approximation (EWA) \cite{muonsmasher,ruiz22}.

\begin{figure}[ht!]
\centering
\includegraphics[scale=0.4]{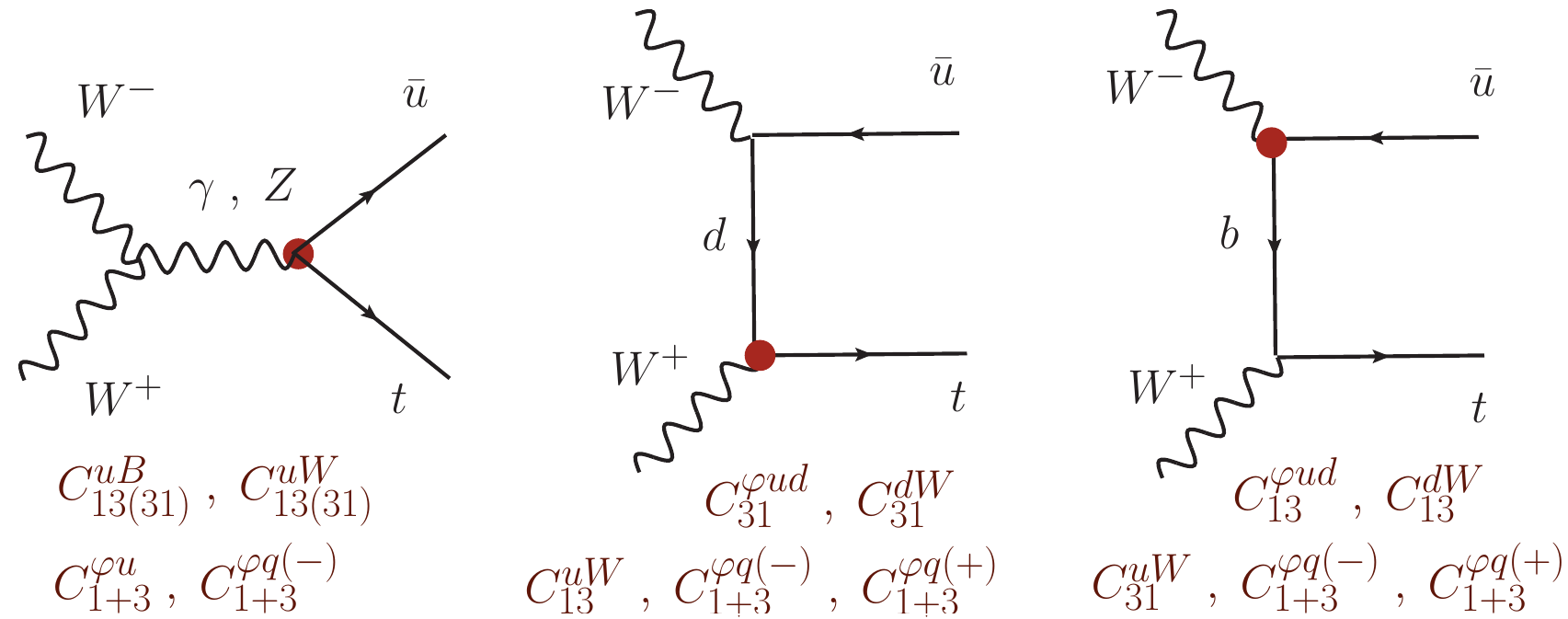}  
\caption{The $WW$ fusion processes
  $W^- W^+ \to t\bar u$: s-channel and t-channel diagrams.}
  \label{fig:wwtu}
\end{figure}

Let us focus on these t-channel $WW$ fusion diagrams without the
leptonic lines as presented in fig.~\ref{fig:wwtu} where the
coefficient associated to each effective coupling is shown for
clarity. We are not making use of the EWA in our calculations but
it is instructive to see what happens when we compute the
two-to-two amplitudes of fig.~\ref{fig:wwtu}.
Regarding the CKM matrix elements, it is known that the
third family subleading flavor mixing elements are very
small, of order $10^{-2}$ or less.
We have performed several of the
calculations keeping all the CKM elements, and then we have made the
assumption $V_{ts}=V_{td}=V_{ub}=V_{cb}=0$ and $V_{tb}=1$. As expected,
we have found that the numerical results are essentially the same.
From now on, we will make our discussion bearing in mind this
(Cabbibo matrix) simplification.
In (\ref{eq:wwtu}) we show the cross section for
$W^+W^- \to t\bar u$ using the contribution of each operator
coefficient separately.  They are given for three sample energies
$\sqs = 1,2,4$ TeV and we are writing them in ascending order:
\bea
\sigma_{ww} /|C^{\vphi ud}_{13}|^2 &=&
40 \; ,\; 29 \; ,\; 27 \;\; {\rm ab}.
\nonumber \\
\sigma_{ww} /|C^{dW}_{13}|^2 &=&
520 \; ,\; 430 \; ,\; 410 \;\; {\rm ab}.
\nonumber \\
\sigma_{ww} /|C^{uB}_{13(31)}|^2 &=&
3.7 \; ,\; 3.4 \; ,\; 3.3 \;\; {\rm fb}.
\nonumber \\
\sigma_{ww} /|C^{u\vphi}_{13(31)}|^2 &=&
75 \; ,\; 77 \; ,\; 78 \;\; {\rm fb}.
\nonumber \\
\sigma_{ww} /|C^{\vphi q(-)}_{1+3}|^2 &=&
400 \; ,\; 330 \; ,\; 310 \;\; {\rm fb}.
\label{eq:wwtu} \\
\sigma_{ww} /|C^{\vphi u}_{1+3}|^2 &=&
1.9 \; ,\; 7.1 \; ,\; 28 \;\; {\rm pb}.
\nonumber \\
\sigma_{ww} /|C^{\vphi q(+)}_{1+3}|^2 &=&
8.0 \; ,\; 28 \; ,\; 110 \;\; {\rm pb}.
\nonumber \\
\sigma_{ww} /|C^{uW}_{13(31)}|^2 &=&
9.0 \; ,\; 29 \; ,\; 112 \;\; {\rm pb}.
\nonumber
\eea

The first line in (\ref{eq:wwtu}) shows the results for operator
$Q_{\vphi ud}$.  It turns out that $C^{\vphi ud}_{31}$ yields a cross section
equal to zero.  This is because of a chirality flip on the $d$-quark
line causing the amplitude to be proportional to $m_d$.
On the other hand, $C^{\vphi ud}_{13}$ refers to a diagram with a $b$-quark
propagator and the amplitude is now proportional to $m_b$ ($4.5$ GeV).
A similar situation occurs with the operator $Q_{dW}$.

The third smallest $WW$ fusion cross section in (\ref{eq:wwtu})
comes from the tensor operator $Q_{uB}$.
This can be regarded as a surprise because the couplings of
tensor operators are proportional to the gauge boson momentum.
Indeed, let us be reminded that the operator with the highest
$\mu^+ \mu^- \to t\bar u$ cross section is precisely $Q_{uB}$.
What happens in $W^+ W^- \to t\bar u$ is that there is a destructive
interference between the two diagrams: the one with a photon
and the one with a $Z$ propagator.  Nevertheless, when we compute
the cross section of the full
$\mu^- \mu^+ \to t\bar u \nu_\mu \bar \nu_\mu$ process we find out
that $Q_{uB}$ actually yields the second highest cross section.
The explanation is that the $WW$ fusion diagrams are not significant
for $Q_{uB}$.  The other diagrams are providing all the total
cross section.
As shown in fig.~\ref{fig:mmtunn}, they can be viewed as $\mu^+ \mu^-$
annihilation into $t\bar u$ plus the emission of a neutrino pair
from any of the fermion lines.

About the inclusion of $Q_{\vphi q(+)}$, we should bear in mind that
this operator generates FCNC $Zbu$ and $Zbs$ couplings
that can be strongly constrained by $B$ meson measurements.
However, a global analysis in the future may take into account the
potential contribution of $Q_{\vphi q(+)}$ to the single top-quark
production process.  After all, in our list of cross sections,
$Q_{\vphi q(+)}$ yields the second highest numbers.
Remarkably, it is in fact almost identical to the yield of $Q_{uW}$.

Concerning the validity of eq.~(\ref{eq:wwtu}) as far as unitarity
violation is concerned, we point out that, the limits on the
coefficients that we obtain in tables \ref{tab:tunnlimhi} and
\ref{tab:tunnlimed} are well below the unitarity limits.
In particular, we find from ref.~\cite{corbett} that the bounds
are given by $|C_Q s/\Lambda^2|< 28,111,14,34$ for
$C_Q=C^{uW}_{13},C^{uB}_{13},C^{\vphi q \pm}_{1+3},C^{\vphi u}_{1+3}$
respectively ($\Lambda =1$TeV, $\sqs = 1,2,4$TeV).

Now let us go ahead with the discussion of the process of interest.
In eq.~(\ref{eq:tunn}) we show the cross section of
$\mu^- \mu^+ \to t\bar u \nu_\mu \bar \nu_\mu$
for each operator coefficient separately.
They are given for the four energies of table~\ref{tab:lumin}
and we are writing them again in ascending order:
\bea
\sigma /|C^{\vphi ud}_{13}|^2 &=&
0.016 \; ,\; 0.032 \; ,\; 0.047 \; ,\; 0.058 \;\; {\rm ab}.
\nonumber \\
\sigma /|C^{dW}_{13}|^2 &=&
0.31 \; ,\; 0.76 \; ,\; 1.3 \; ,\; 1.8 \;\; {\rm ab}.
\nonumber \\
\sigma /|C^{u\vphi}_{13(31)}|^2 &=&
18 \; ,\; 38 \; ,\; 58 \; ,\; 74 \;\; {\rm ab}.
\nonumber \\
\sigma /|C^{\vphi q(-)}_{1+3}|^2 &=&
0.09 \; ,\; 0.19 \; ,\; 0.29 \; ,\; 0.37 \;\; {\rm fb}.
\label{eq:tunn} \\
\sigma /|C^{\vphi u}_{1+3}|^2 &=&
0.19 \; ,\; 0.86 \; ,\; 2.5 \; ,\; 5.0 \;\; {\rm fb}.
\nonumber \\
\sigma /|C^{\vphi q(+)}_{1+3}|^2 &=&
0.67 \; ,\; 3.0 \; ,\; 8.8 \; ,\; 17 \;\; {\rm fb}.
\nonumber \\
\sigma /|C^{uB}_{13(31)}|^2 &=&
1.2 \; ,\; 4.6 \; ,\; 13 \; ,\; 25 \;\; {\rm fb}.
\nonumber \\
\sigma /|C^{uW}_{13(31)}|^2 &=&
4.0 \; ,\; 21 \; ,\; 69 \; ,\; 147 \;\; {\rm fb}.
\nonumber
\eea
The operators $Q_{\vphi ud}$ and $Q_{dW}$ yield remarkably small cross
sections, showing that these are indeed low-sensitivity
operators. Moreover, their coefficients get very strong limits, of
order less than $10^{-2}$, from $B$ meson decays and $pp \to \ell \nu$
at the LHC \cite{isidori2010,greljo23}. These operators can therefore
be ignored in processes based on top production at the MuC.

In the third place we have $Q_{u\vphi}$.  This operator belongs to the group
that only generates couplings with the top quark, so we do not expect
any strong constraints from $B$ measurements.  The current limits from
non observation of $t\to uH$ and $t\to cH$ decays at the LHC as reported
by ATLAS are $|C_{13(31)}^{u\vphi}| \leq 0.96$ and
$|C_{23(32)}^{u\vphi}| \leq 1.1$ \cite{atlastqh23}.
In principle, we can imagine that Higgs associated production
$\mu^+ \mu^- \to t {\bar u}H$ may be a good probe for this coupling.
However, we have obtained cross sections of order $10^{-1}$ ab or less
at all energies (with coefficient equal to 1).
For $t\bar u \nu_\mu \bar \nu_\mu$ production we obtain not as small
cross section numbers, but they may not be enough to get limits
of order 1.  Therefore, the sensitivity for $C_{13(31)}^{u\vphi}$
at the MuC is likely to be somewhat weaker
than the current constraints from the LHC.
In any case, associated production of top and Higgs may still be
an important measurement at the MuC, in a recent article the authors
show that CP violating effects of the top Yukawa coupling $ttH$ may be
probed through $t\bar t H$ production mode \cite{Cassidy:2023lwd}.

Regarding the other six operators, when computing
the amplitude for $\mu^- \mu^+ \to t\bar u \nu_\mu \bar \nu_\mu$
it turns out that one gets many more possible diagrams than in
the associated $t\bar u \gamma (Z)$ production.  Specifically,
for $Q_{\vphi u}$ and $Q_{\vphi q(-)}$ we obtain $15$ diagrams.
For $Q_{uB}$ we obtain $22$ and for $Q_{uW}$ there are $26$.
When we consider $Q_{\vphi u}$ or $Q_{\vphi q(-)}$ the s-channel $WW$
fusion diagrams only involve the effective $tuZ$ vector coupling.

As noted in the previous section $Q_{\vphi q(+)}$ and $Q_{\vphi q(-)}$
generate a $Wub$ coupling which can be probed with the $b\to u \ell \nu$
decay for example.  In fact, very strong limits: of order
$2\times 10^{-3}$, have been reported in a recent article \cite{greljo23}.
In addition, there are also much weaker limits from the non-observation
of $t\to uZ (cZ)$ decay at the LHC as reported by ATLAS \cite{atlas2023}:
$|C_{1+3}^{\vphi q(-)}| \leq 0.10$ and $|C_{2+3}^{\vphi q(-)}| \leq
0.14$ (see appendix \ref{sec:lhc.bounds}).
We point out that similar limits on $|C_{a+3}^{\vphi q(-)}|$ can be found in
ref.~\cite{durieux}, where they also used $tj$ production at LEP data.
A surprising result in eq.~(\ref{eq:tunn}) is that it is actually
$Q_{\vphi q(+)}$, not $Q_{\vphi q(-)}$, the operator that gives a greater
cross section for $t {\bar u} \nu_\mu \bar \nu_\mu$ production.
In any case, we should bear in mind that both of them receive very
strong limits from $B$ measurements.

The fact that, except for $Q_{uB}$, the same ranking that we observe
in eq.~(\ref{eq:wwtu}) appears again in eq.~(\ref{eq:tunn}) points
toward the significance of $WW$ fusion in the production processes of
a high energy MuC.  It is because of this feature of a high energy
MuC, that it has been dubbed as a virtual {\it gauge boson collider}
\cite{muonsmasher}.  However, and contrary to this statement, there
may be processes where the $WW$ fusion diagrams are not so significant
even at high energies.  The reason why $Q_{uB}$ yields such high rates
in the full process is because there are $20$ other diagrams, like the
ones on the right-side of of fig.~\ref{fig:mmtunn} that involve FCNC
$\gamma tu$ and $Ztu$ couplings.  They grow with energy and there is
apparently no cancellation taking place between them.  None of them
correspond to a gauge boson fusion process. As a confirmation that our
reasoning is correct we will now turn our attention to another
$t\bar u$ plus neutrinos production process: one where the neutrinos
are not the $SU(2)$ partners of the muon.  In this process there
cannot be any $WW$ fusion diagram.  It turns out that the cross
section for $Q_{uB}$ when we substitute $\nu_\mu$ for $\nu_e$ or
$\nu_\tau$ is almost exactly the same. The same does not happen with
$Q_{uW}$, for which the contribution from the other neutrino final
states is small.  Below, we give the cross section for these additional
modes: 
\begin{table}[ht!]
  \centering
  \begin{tabular}{|cc|cccc|}\hline
    $\sigma (fb)$ & ${\;}$ & $3$TeV & $6$TeV & $10$TeV & $14$TeV
 \\\hline
 $C^{uB}_{13(31)}$ & $\nu_e + \nu_\tau$ &  
  $2.3$ & $9.3$ & $26$ & $51$
\\
  ${\;}$ & $\nu_e + \nu_\mu + \nu_\tau$ & 
$3.5$ & $14$ & $39$ & $76$
\\\hline
 $C^{uW}_{13(31)}$ & $\nu_e + \nu_\tau$ &  
    $1.5$ & $6.1$ & $17$ & $33$
\\
 ${\;}$ & $\nu_e + \nu_\mu + \nu_\tau$ & 
  $5.6$ & $27$ & $75$ & $180$
\\\hline
  \end{tabular}                               
  \caption{Cross sections
    $\sigma (\mu^- \mu^+ \to t\bar q \nu \bar \nu)(fb)$
    for the modes $\nu = \nu_e , \nu_\tau$. 
    The sum of the three modes $\nu = \nu_e , \nu_\mu ,\nu_\tau$ is
  also shown.}
  \label{tab:tunne}
\end{table}
With the total cross sections shown in table~\ref{tab:tunne}
we now make an estimate on the sensitivity
from $\mu^- \mu^+ \to t\bar u \nu \bar \nu$ at the MuC
for the operators of our main interest in this study.  We 
are assuming that $5\%$ of the signal events will pass the cuts
and request a value of the coefficient necessary for $25$ events.
In table~\ref{tab:tunnlimhi} we show the limits for
$C^{uB}_{13(31)}$ and $C^{uW}_{13(31)}$. We also show
the ratio with respect to the current ATLAS limits presented in
table~\ref{tab:hilimits}.  We notice that at the highest energies
the MuC may provide bounds two to four times more
stringent than what the LHC has obtained from the most sensitive
probe that is $t\to u\gamma$ decay.
In the following section we will perform a more in-depth
study of the limits for $C^{uW,uB}_{13,31}$ by
considering the decay products of top quark in the hadronic
channel.

\begin{table}[ht!]
  \centering
  \begin{tabular}{|c|c|c|c|c|}\hline
    ${\;\;}$  & $3$TeV & $6$TeV & $10$TeV & $14$TeV
    \\\hline
  $10^2 \times |C^{uB}_{13(31)}|\leq$ &
  $54{\;}(R=14)$ & $13{\;}(R=3.4)$ &
  $5.1{\;}(R=1.3)$ & $2.6{\;}(R=0.65)$ 
  \\\hline
  $10^2 \times |C^{uW}_{13(31)}|\leq$ &
  $42{\;}(R=6.0)$ & $9.6{\;}(R=1.4)$ &
  $3.7{\;}(R=0.53)$ & $1.7{\;}(R=0.24)$
\\\hline
  \end{tabular}                               
  \caption{Limits on the coefficients from
    $\mu^- \mu^+ \to t\bar u \nu \bar \nu$ obtained with the
minimal cross sections required for $1000$ $t\bar u$ events.
The ratios
$R=|C^{\rm max}_{\mu {\rm MuC}}| / |C^{\rm max}_{\rm LHC}|$ are shown.}
  \label{tab:tunnlimhi}
\end{table}


To end this section, in table~\ref{tab:tunnlimed} we also show the limits
obtained for the other couplings.  For $C^{u \vphi}_{13(31)}$
we notice that the MuC could only provide similar bounds to the
FCNC top-Higgs couplings that the LHC has already achieved. Concerning
$C^{\vphi q(-)}_{13(31)}$ the current LHC bounds from $t\to uZ$ are $0.1$
which is already twice as strong as the lowest limit in
table~\ref{tab:tunnlimed}.  We also show limits for
$C^{\vphi q(+)}_{13(31)}$ even though an FC bottom quark production would
likely be the best process to use as probe of this operator.
In any case, we think it is worth noting that even single top
production could be a somewhat sensitive measurement.
To underline the importance of the limits shown in
table~\ref{tab:tunnlimed} we point out that there are models
that give rise to tree level contributions to vector operators
like $Q_{\vphi q (\pm)}$ and $Q_{\vphi u}$ but only loop level
contributions to the tensor operators in table~\ref{tab:tunnlimhi}
\cite{Crivellin:2022fdf}.

\begin{table}[ht!]
  \centering
  \begin{tabular}{|c|c|c|c|c|}\hline
    ${\;\;}$  & $3$TeV & $6$TeV & $10$TeV & $14$TeV
    \\\hline
  $|C^{u \vphi}_{13(31)}|\leq$ &
    $7.5{\;}(R=6.2)$ & $2.6{\;}(R=2.2)$ &
    $1.3{\;}(R=1.1)$ & $0.82{\;}(R=0.69)$
    \\\hline
   $|C^{\vphi u}_{1+3}|\leq$ &
    $2.3{\;}(R=12)$ & $0.54{\;}(R=2.8)$ &
    $0.20{\;}(R=1.1)$ & $0.10{\;}(R=0.53)$
  \\\hline  
  $|C^{\vphi q(-)}_{1+3}|\leq$ &
  $3.3{\;}(R=33)$ & $1.2{\;}(R=12)$ &
  $0.59{\;}(R=5.9)$ & $0.37{\;}(R=3.7)$
  \\\hline
  $|C^{\vphi q(+)}_{1+3}|\leq$ &
  $1.2{\;}(R=600)$ & $0.29{\;}(R=145)$ & $0.11{\;}(R=55)$ & $0.054{\;}(R=27)$
\\\hline
  \end{tabular}                               
  \caption{Limits on the coefficients from
    $\mu^- \mu^+ \to t\bar u \nu_\mu \bar \nu_\mu$ obtained with the
    minimal cross sections required for $1000$ $t\bar u$ events.
    For the ratio $R$ of $C^{\vphi q(+)}_{1+3}$ we use the limit of
    $2\times 10^{-3}$ from $B$-meson decays \cite{greljo23}. }
  \label{tab:tunnlimed}
\end{table}


\section{Limits on
  $\boldsymbol{|C^{uW}_{k3,3k}|}$ and $\boldsymbol{|C^{uB}_{k3,3k}|}$
  from $\boldsymbol{t\bar{q}_u\nu\bar{\nu}}$ production 
  in hadronic channel.}
\label{sec:hadr.chan}

In this section we consider the single-top production process
discussed in section \ref{sec:mmtunn} in the hadronic decay channel
which, at the parton level, leads to six-fermion final states,
\begin{equation}
  \label{eq:hc.1}
 \mu^+\mu^- \rightarrow t \bar{q}_u \nu \bar{\nu} + \bar{t} q_u \nu
 \bar{\nu} \rightarrow
 b q_u \bar{q}_d \bar{q}_u \nu \bar{\nu} +
 \bar{b} \bar{q}_u q_d q_u \nu \bar{\nu},
 \quad
 \mathrm{with} \;
 q_u = u, c, \; q_d = d, s, \;
 \nu = \nu_e, \nu_\mu, \nu_\tau.
\end{equation}
This equation defines our parton-level signal process, which is not
allowed in the SM at tree level and must proceed, therefore, through
effective FC couplings. 
We simulate the signal process and its SM
backgrounds in the unitary-gauge SM, with two massless generations,
with Cabibbo mixing in the quark sector, augmented by the dimension-6
operators (\ref{eq:operators}). As mentioned above, we implement those
operators in MG5 \cite{alw14} with FeynRules \cite{all14}, and analyze
the MG5 events with ROOT \cite{root}.

Representative Feynman diagrams for (\ref{eq:hc.1}) can be obtained by
attaching appropriate hadronic decay vertices to the top lines in
figure (\ref{fig:mmtunn}).  We let
$C^{uW}_{13}\neq 0 \neq C^{uW}_{31}$ and
$C^{uB}_{13}\neq 0 \neq C^{uB}_{31}$ one at a time, and set all other
Wilson couplings to vanish. For each Wilson coefficient there are
diagrams containing $n$ effective vertices, with $1\leq n\leq4$. In
practice, however, for $|C^{uW,uB}_{ij}| \lesssim 1$, only diagrams
with one anomalous vertex are numerically significant. Those diagrams
involve the off-diagonal couplings $C^{uW,uB}_{k3,3k}$ with $k=1$,
2. In the remainder of this section we refer to $C^{uW,uB}_{13,31}$
for concreteness, in the understanding that the results obtained apply
equally well to $C^{uW,uB}_{23,32}$.

\begin{figure}[ht!]
\centering
\includegraphics[scale=0.4]{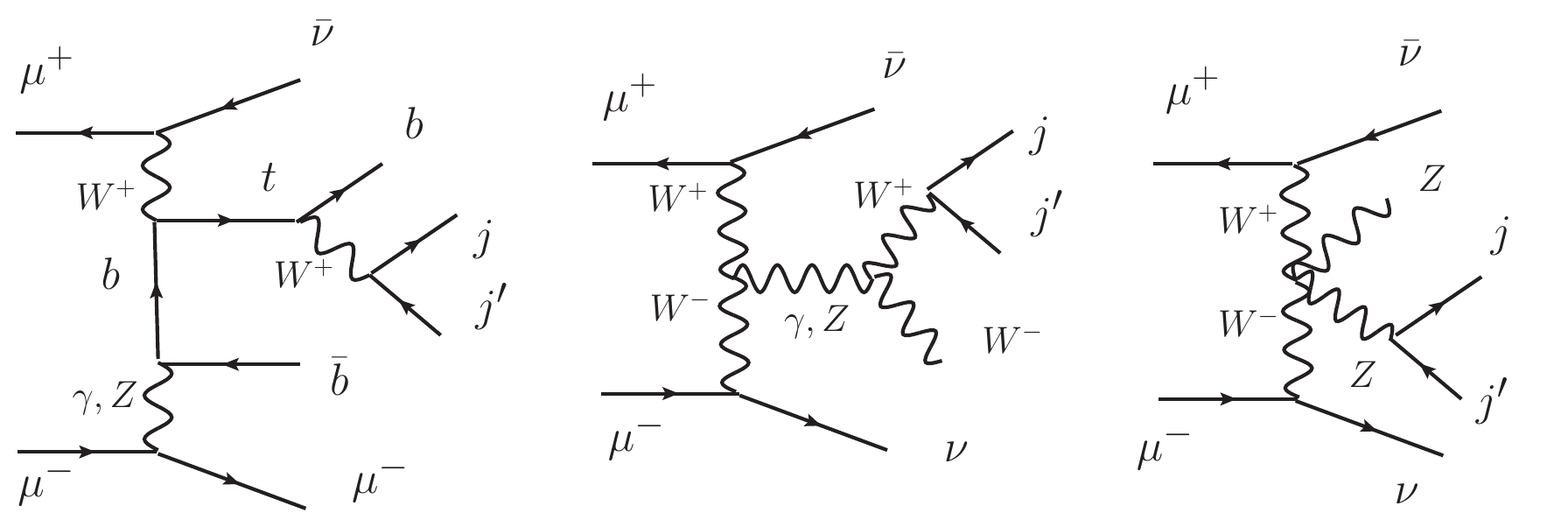}  
\caption{The background processes: single top, $Zjj$ and $Wjj$
 production, some sample diagrams.}
  \label{fig:mmtbnn}
\end{figure}

Besides the signal process (\ref{eq:hc.1}), we take into account
several SM backgrounds with four partons and one or more neutrinos in
the final state. In what follows, we focus on the most resonant
processes.  SM $t\bar{t}$ production through $\mu^+\mu^-$
annihilation, with one top decaying hadronically according to the SM,
and the other via an effective FC coupling 
as $t\rightarrow q_u Z\rightarrow q_u \nu\bar{\nu}$, leads to the same
final states as in (\ref{eq:hc.1}). The cross section for this
process, however, is negligibly small compared to that of the signal
process, so we will not dwell further on it. 
Similarly, we consider triply resonant $W^+W^-Z$ and $ZZZ$ production,
with two bosons decaying hadronicall and one
$Z\rightarrow\nu\bar{\nu}$. These processes also have negligibly small
cross sections, so we ignore them in what follows. 

We considered also doubly-resonant production of $WW$, $WZ$ and $ZZ$
vector bosons. Out of those, only the first one yields a large enough
cross section to merit further discussion. A Feynman diagram for the
production of $WW$ in a vector-boson fusion (henceforth VBF) process
is shown in figure \ref{fig:mmtbnn}. Other topologies, such as
$\mu^+\mu^-$ annihilation, are not shown for brevity.
This $WW$ background can be associated with any of the three neutrino
flavors, but it is only the process with $\nu=\nu_\mu$,
which is the only one containing the VBF diagrams,
that gives a numerically significant contribution. As can be seen
in the figure, this is a reducible background, with no $b$ quark in
the final state. Numerical results for this $WW$ background cross
section are given below in table \ref{tab:sigma} under the label
``$WW$.'' 

Singly resonant $Wjj$ production is another important reducible
background process. A Feynman diagram for this process is given by the
$WW$ diagram in figure \ref{fig:mmtbnn}, with one of the $W$ bosons
on-shell and the other one off-shell. As with $WW$ production, only the
diagrams with $\nu=\nu_\mu$ of the VBF type are numerically relevant.
Numerical results for the cross sections for this background are given
below in table \ref{tab:sigma} with the label ``$Wjj$.''
Another important singly-resonant background is $Zjj$ production.
The diagrams containing the decay $Z\rightarrow q\bar{q}$, with
$q$ any quark flavor including $q=b$, and $\nu=\nu_\mu$ are the
diagrams that give the only numerically relevant contribution to
the single-$Z$ production cross section.
Numerical results for the cross section for this background are
given below in table \ref{tab:sigma} under the label ``$Zjj$.''

We consider, finally, SM top production in $\mu^+\mu^-$ collisions.
Top-pair production from $\mu^+\mu^-$ annihilation, in
semileptonic mode, with the charged lepton outside the central region,
is a reducible background. Its cross section is very small, however,
so we do not consider it further. Top-pair production of the form
$\mu^+\mu^- \rightarrow t \bar{t} \nu \bar{\nu}$, which contains
$t\bar{t}$ production in VBF among other processes, is also a
reducible background in semileptonic mode with the charged lepton
outside the central region. Its cross section is larger than that of
the annihilation process, but still negligibly small.  The main SM
top-production background at high energies is single-top production in
association with a $b$ jet,
\begin{equation}
  \label{eq:hc.2}
  \mu^+\mu^- \rightarrow t \bar{b} \ell^- \nu_\ell +
  \bar{t} b \ell ^+ \bar{\nu}_\ell
  \rightarrow
  bq_u\bar{q}_d\; \bar{b} \ell^- \nu_\ell +
  \bar{b} \bar{q}_u q_d\; b \ell ^+ \bar{\nu}_\ell.
\end{equation}
This process is given by $t$ (or $\bar{t}$) production with $\ell$
any of the three charged leptons, but again only the processes
with $\ell=\mu$ lead to a substantial cross section.  Among those,
only the VBF processes are significant.  It is illustrated by the
leftmost diagram in figure \ref{fig:mmtbnn}.
Single-top production has two $b$ quarks in the final state, and is
therefore a reducible background. Numerical results for its cross
section are given below in table \ref{tab:sigma} with the
label ``$tb$.''

\begin{figure}[ht]
  \centering{}
  \includegraphics[scale=0.9]{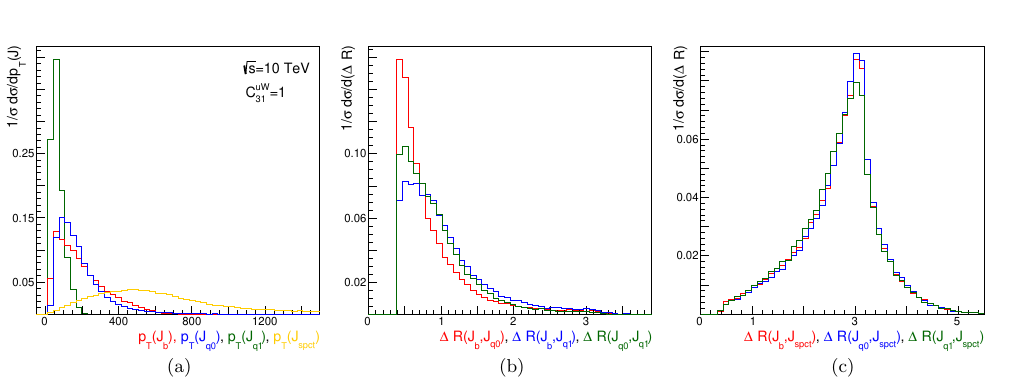}
  \caption{Differential cross sections for the signal process
    (\ref{eq:hc.1}), at $\sqrt{s}=10$ TeV and for $C^{uW}_{31}=1$ and
    all other effective FC couplings 
    vanishing, with respect to (a) the
    transverse momentum of the top decay products $J_b$, $J_{q0}$,
    $J_{q1}$, and the spectator jet $J_\mathrm{spct}$, (b) the
    $\Delta R$ distance between top decay products, (c) the $\Delta R$
    distance between top decay products and the spectator jet.}
  \label{fig:dsigma}
\end{figure}

We carry out the event selection of our simulated signal and
background events by means of a set of phase-space cuts. We consider
final states with four partons, which we interpret as ``jets,'' and
transverse missing energy. We denote the jets $J_{0,\ldots,3}$,
ordered by decreasing $p_T$. We then select events satisfying the
phase-space cuts,
\begin{equation}
  \label{eq:hc.3}
  \begin{gathered}
    p_T(J_0)\geq \ldots \geq p_T(J_3) > 30\,\mathrm{GeV},
    \quad
    \not\!\!E_T > 50\, \mathrm{GeV},
    \quad
    |y(J_{0,\ldots,3})| < 3,\\
    \Delta R(J_a,J_b) > 0.4,
    \quad
    m(J_a,J_b) > 20\,\mathrm{GeV},
    \quad
    a\neq b=0,\ldots,3,
  \end{gathered}
\end{equation}
where $p_T=|\vec{p}^\perp|$ is the transverse momentum,
$\not\!\!E_T=|\vec{p}^\perp_\nu+\vec{p}^\perp_{\bar{\nu}}|$ is the
missing transverse energy, and
$\Delta R=\sqrt{(\Delta y)^2 +(\Delta \varphi)^2 }$.  The purpose of
the cuts (\ref{eq:hc.3}) is fourfold: (i) to ensure that the jets are
in the central region where $b$-tagging is operational and efficient,
(ii) to ensure good numerical convergence of the simulation
algorithms, (iii) to suppress background processes, whose jets tend to
be somewhat softer than those in the signal processes, and (iv) to
select events with well separated partons in the final state, which
are the parton-level idealization of four-jet events at the detector
level.  This last point deserves further consideration. At the
extremely high energies the muon collider is foreseen to reach, we can
expect events with highly boosted top quarks resulting in fewer and
fatter jets in the final state. This is indeed the case in two-body
reactions of the form $\mu^+\mu^-\rightarrow t\bar{q}_u+\bar{t}q_u$
\cite{sun2023}. For processes with additional leptons in the final
state, and in particular for VBF processes, however, a significant
portion of the cross section is provided by four-jet events. This is
illustrated by the differential cross sections shown in figure
\ref{fig:dsigma}. In this preliminary study, we focus our analysis
precisely on those four-jet events, which are enough to attain
substantial sensitivity to the effective flavor off-diagonal
couplings $C^{uW}_{13,31}$, and $C^{uB}_{13,31}$, as shown below.

Each jet is assigned a $b$-tag, for which purpose we adopt a working
point with a $b$-tagging efficiency $\eta_b=0.85$, and mistagging
probabilities $p_c=0.10$ for $c$-jets and $p_j=0.01$ for lighter jets.
Our event selection is further refined by requiring the same number of
$b$-tagged and light jets as in (\ref{eq:hc.1}),
\begin{equation}
  \label{eq:hc.4}
  N_b =1, \quad N_j=3.
\end{equation}
For events satisfying (\ref{eq:hc.4}), we designate the $b$-tagged jet
as $J_b$. We construct the mass differences
$\Delta m = |m(j,j')-m_W|$, for non-$b$ tagged jets $j$, $j'$, and
identify the pair of jets minimizing $\Delta m$ as the $W$ decay
products. We denote those jets as $J_{q0}$, $J_{q1}$, ordered by
$p_T$. The remaining light jet is identified as the light ``spectator''
jet produced in the effective FC coupling, 
and denoted $J_\mathrm{spct}$. We then require the events to satisfy
the mass cuts
\begin{equation}
  \label{eq:hc.5}
  |m(J_{q0},J_{q1})-m_W| < 30\,\mathrm{GeV},
  \quad
  |m(J_b,J_{q0},J_{q1})-m_t| < 30\,\mathrm{GeV},
\end{equation}
where $m_W=80.377$ GeV, $m_t=172.69$ GeV \cite{PDG}.
In figure \ref{fig:dsigma} we show the distribution of the transverse
momentum of, and the phase-space distance between, the four
final-state jets.  As seen there, there is a substantial fraction of
events where the final-state particles are well separated.
Finally, in the case of SM top production, as discussed above, we
require the final-state charged lepton to be outside the central
region,
\begin{equation}
  \label{eq:hc.6}
  \mathrm{SM:} \quad |y(\ell^\pm)| > 3.
\end{equation}

With the set of cuts described above, we obtain the cross sections for
the signal and background processes. The results are summarized in
table \ref{tab:sigma}, where the cross sections for the signal process
(\ref{eq:hc.1}) are given for the
effective FC couplings 
$C^{uW}_{13,31}$ and $C^{uB}_{13,31}$ taken non-zero one at a time.

As expected by the total cross sections in eq.~(\ref{eq:tunn}), we
observe here that the signal cross sections for $C^{uW}_{13,31}\neq0$,
with cuts, grow with $\sqrt{s}$.  This reflects both, the dominating
contribution of VBF for these couplings, and the fact that a significant
fraction of events contains well separated final-state particles
satisfying (\ref{eq:hc.3}).
For $C^{uB}_{13,31}$, we notice in eq.~(\ref{eq:tunn}) that the growth
in energy is slower and then, after imposing the cuts, the growth
disappears.  This reflects the fact that the VBF process is not
dominant for this coupling.

Also important are the results shown in table
\ref{tab:sigma} for the main background processes, as described above.

\begin{table}[h]
  \centering{}
  \begin{tabular}{ccccc}
    &\multicolumn{4}{c}{$\sigma$ [fb]}\\\cline{2-5}
    $\sqrt{s}$ [TeV]&3&6&10&14\\\hline
    $\sigma_{d6}/|C^{uW}_{13}|^2$&1.254&2.568&3.709&4.446\\
    $\sigma_{d6}/|C^{uW}_{31}|^2$&1.382&3.097&4.687&5.716\\
    $\sigma_{d6}/|C^{uB}_{13}|^2$&0.418&0.412&0.364&0.308\\
    $\sigma_{d6}/|C^{uB}_{31}|^2$&0.469&0.485&0.451&0.376\\    
    $WW$                    &0.263&0.371&0.454&0.486\\
    $Wjj$                   &0.00773&0.0190&0.0184&0.0205\\
    $Zjj$                   &0.302&0.425&0.554&0.516\\
    $tb$                    &0.122&0.282&0.471&0.485\\
    SM                      &0.694&1.065&1.497&1.508
  \end{tabular}
  \caption{Cross sections in fb for the signal process (\ref{eq:hc.1})
    with $C^{uW,uB}_{13,31}=1$ one at a time, and all other
effective FC couplings
    vanishing, and for SM background processes: doubly-resonant
  $W$ production (labeled $WW$), singly-resonant $W$ ($Wjj$) and $Z$
  ($Zjj$) productions, and single-top production ($tb$). The sum of
  all backgrounds is labeled SM. All cross sections computed with the
  cuts (\ref{eq:hc.3})--(\ref{eq:hc.6}).}
  \label{tab:sigma}
\end{table}

We obtain limits on the effective FC couplings 
$|C^{uW}_{13,31}|$ and $|C^{uB}_{13,31}|$ from the relation,
\begin{equation}
  \label{eq:hc.7}
  N_{d6} < \mathcal{S} \Delta_\mathrm{tot}N_\mathrm{SM},
\end{equation}
where $\mathcal{S}$ is the statistical significance, and
\begin{equation}
  \label{eq:hc.8}  
  \begin{gathered}
  N_{d6} = \sigma_{d6} L_\mathrm{int} = |C^{uW,uB}_{13,31}|^2
  \left(\frac{\sigma_{d6}}{|C^{uW,uB}_{13,31}|^2}\right) L_\mathrm{int},
  \quad
  N_\mathrm{SM} = \sigma_\mathrm{SM} L_\mathrm{int}, \\
  \Delta_\mathrm{tot}N_\mathrm{SM} =
  \sqrt{(\Delta_\mathrm{stat}N_\mathrm{SM})^2 +
    (\Delta_\mathrm{syst}N_\mathrm{SM})^2},
  \quad
  \Delta_\mathrm{stat}N_\mathrm{SM} = \sqrt{\sigma_\mathrm{SM}
  L_\mathrm{int}},
  \quad
  \Delta_\mathrm{syst}N_\mathrm{SM} = \varepsilon_\mathrm{syst} \sigma_\mathrm{SM}
  L_\mathrm{int},
  \end{gathered}
\end{equation}
with $(\sigma_{d6}/|C^{uW,uB}_{13,31}|^2)$ taken from the first two lines of
table \ref{tab:sigma}, $\sigma_\mathrm{SM}$ from the last line of that
table, and the integrated luminosity $L_\mathrm{int}$ from table
\ref{tab:lumin}. The relative systematic uncertainty
$\varepsilon_\mathrm{syst}$, which must be experimentally established,
is tentatively set in this paper to 2\%,
$\varepsilon_\mathrm{syst}=0.02$. The limits on $|C^{uW,uB}_{k3,3k}|$,
$k=1$, 2, obtained from (\ref{eq:hc.7}) are shown in table
\ref{tab:limit} for four values of $\sqrt{s}$. As expected, in view of
the increasing cross sections in table \ref{tab:sigma}, the
sensitivity is higher the larger the collision energy. Also shown in
the table are limits obtained for purely statistical uncertainty,
$\varepsilon_\mathrm{syst}=0$, to assess the relative importance of
statistical and systematic uncertainties.

\begin{table}
  \centering{}
  \begin{tabular}{ccccccccccc}
$\sqrt{s}$ [TeV]  &3&6&10&14&\rule{0.5cm}{0pt}&$\sqrt{s}$ [TeV]&3&6&10&14\\\cline{2-5}\cline{8-11}        
& \multicolumn{4}{c}{$\varepsilon_\mathrm{syst}=0.00$}
&\rule{0.5cm}{0pt}&&\multicolumn{4}{c}{$\varepsilon_\mathrm{syst}=0.00$}\\\cline{2-5}\cline{8-11}        
$|C^{uW}_{13}|<$&0.145&0.0797&0.0574&0.0442&\rule{0.5cm}{0pt}&$|C^{uB}_{13}|<$&0.251&0.199&0.183&0.168\\
$|C^{uW}_{31}|<$&0.138&0.0276&0.0511&0.0390&\rule{0.5cm}{0pt}&$|C^{uB}_{31}|<$&0.237&0.183&0.165&0.152\\\cline{2-5}\cline{8-11}        
& \multicolumn{4}{c}{$\varepsilon_\mathrm{syst}=0.02$}
&\rule{0.5cm}{0pt}&&\multicolumn{4}{c}{$\varepsilon_\mathrm{syst}=0.02$}\\\cline{2-5}\cline{8-11}
$|C^{uW}_{13}|<$&0.154&0.102 &0.0934&0.0840&\rule{0.5cm}{0pt}&$|C^{uB}_{13}|<$&0.267&0.255&0.298&0.319\\
$|C^{uW}_{31}|<$&0.147&0.0931&0.0831&0.0741&\rule{0.5cm}{0pt}&$|C^{uB}_{31}|<$&0.252&0.235&0.268&0.289    
  \end{tabular}
  \caption{Single-coupling limits on $|C^{uW,uB}_{13,31}|$ at
    statistical significance $\mathcal{S}=1$ (68\% C.L.) from
    (\ref{eq:hc.7}) for systematic uncertainty
    $\varepsilon_\mathrm{syst}=2\%$. Limits for
    $\varepsilon_\mathrm{syst}=0\%$ are also shown for comparison.}
  \label{tab:limit}
\end{table}

\section{Conclusions}
\label{sec:conclusions}

\newcounter{aux}                                                                                                                       
\renewcommand{\theaux}{(\arabic{aux})}

Based on the processes of FC single top-quark production at a high
energy MuC, we have estimated individual limits on dimension-six
operators of the SMEFT basis \cite{grz10}.  In particular, we have
identified a comprehensive list of nine operators that generate top
quark-gauge boson couplings and seven four-fermion operators for
the contact terms $\mu \mu tu$. Based on the potential yields on the
cross sections and the current limits in the literature we have found
that the MuC can have widely different sensitivities to these
operators.  Operators $Q_{\vphi ud}$ and $Q_{dW}$ get very weak
constraints of $O(10^1)$, whereas $Q_{uW}$ and $Q_{uB}$ get limits
of $O(10^{-2})$.  Four-fermion operators get limits of $O(10^{-4})$
while current limits are of order 1.  Below we present a detailed
list of the specific conclusions of our study.

\begin{trivlist}
  \refstepcounter{aux}
\item\theaux\label{aaa} There are seven four-fermion operators in the
  Warsaw basis that contribute to $t\bar u$ (and $u\bar t$)
  production.  We define two combinations that separate $\mu \mu tu$
  and $\mu \mu bd$ contact terms so we reduce the list to six
  operators.  Since the cross sections grow with $s$, the limits we
  obtain are three to four orders of magnitude smaller than the
  current limits from LEP and LHC data.

  \refstepcounter{aux}  
\item[ \theaux\label{zero}] Concerning fermion-boson operators, two of
  them stand out as the most relevant for FC top production at the MuC:
  $Q_{uB}$ and $Q_{uW}$.  At tree level they are not
  sensitive to $B$ measurements nor FC bottom quark production.
  For energies $6-14$ TeV the MuC would give better limits to these
  two operators than the current 2023 LHC bounds (see table
  \ref{tab:hilimits}), though the future HL-LHC might yield similar
  limits.  The two most important processes are
  $\mu^+ \mu^- \to t \bar q$ and
  $\mu^+ \mu^- \to t \bar q \nu \bar \nu$, with $\nu$ representing the
  contribution from the three neutrino flavors. The limits from these
  two processes are given in tables \ref{tab:tulimits} and
  \ref{tab:tunnlimhi}, respectively (see also table
  \ref{tab:tuzlimits}, for limits from radiative top production).

\refstepcounter{aux}  
\item[ \theaux] $C^{uB}_{k3}$ and $C^{uB}_{3k}$ get better limits from
  the two-to-two process $\mu^+ \mu^- \to t \bar{q} + \bar{t}q$ than
  from $\mu^+ \mu^- \to t \bar{q}\nu\bar{\nu} + \bar{t}q\nu\bar{\nu}$,
  see table \ref{tab:tulimits}. For the latter process with
  $\nu=\nu_\mu$, a strong destructive interference occurs between the
  $WW$ fusion diagrams with $\gamma$ and $Z$ propagators.  Therefore,
  $WW$ fusion only gives a very small contribution and we have found
  that for this operator in particular the EWA does not apply.
  This would not be an isolated case where the approximation differs 
  from the full calculation. Some years ago a study on the validity of
  the EWA for (SM) $e^+ e^- \to W^+ W^- \nu_e \bar \nu_e$ at $\sqs=2$TeV
  found substantial deviations in some kinematical regions
  \cite{Bernreuther2015}.
  
\refstepcounter{aux}  
\item[  \theaux]
  For the case of $\mu^+ \mu^- \to t \bar q \nu \bar \nu$,  
  $C^{uW}_{k3}$ and $C^{uW}_{3k}$ ($k=1,2$) are the coefficients
  that get the strongest limits.  For a MuC at $10$ TeV and
  $10 {\rm ab}^{-1}$ integrated luminosity they can be one third lower
  than the current limits from $t\to u\gamma$ at the LHC. See table
  \ref{tab:limit} for limits from the hadronic decay channel of
  associate top prodution $\mu^+ \mu^- \to t \bar q \nu \bar \nu$.  

  \refstepcounter{aux}  
\item[ \theaux] $C^{\vphi u}_{k+3}$ gets competitive
  limits only from the $WW$ fusion diagram in
  $\mu^+ \mu^- \to t \bar q \nu_\mu \bar \nu_\mu$ and not from
  $\mu^+ \mu^- \to t \bar q$. See table \ref{tab:tunnlimed} in section
  \ref{sec:mmtunn}. Notice, moreover, that unlike the other operators
  in that table, $Q_{\vphi u}^{k+3}$ does not contribute to $B$ decays
  at tree level.

\refstepcounter{aux}  
\item[ \theaux] Concerning $Q_{\vphi q(-)}$, the MuC would only
  provide limits that are already twice as large as the current LHC
  limits that come from $t\to uZ$ decay.  In addition to this,
  $Q_{\vphi q(-)}$ also generates a left handed $Wub$ coupling that is
  sensitive to $B$ meson measurements at tree level.  (See table
  \ref{tab:tunnlimed} for our constraints on this coupling from
  $\mu^+ \mu^- \to t \bar q \nu\bar{\nu}$ production.) 

\refstepcounter{aux}  
\item[ \theaux] Clearly, operators that generate FCNC $Zbd$ and
  $\gamma bd$ couplings should be probed by single bottom quark
  production rather than top production. However, we point out that
  $C^{\vphi q(+)}_{k+3}$ in particular might also be probed
  succesfully even with $t\bar q \nu_\mu \bar \nu_\mu$ production (see
  table \ref{tab:tunnlimed}).

\refstepcounter{aux}
\item[ \theaux] As observed in section \ref{sec:tua.tuz},
  $C^{uG}_{k3(3k)}$ could be probed with the $t\bar u g$ production,
  but the sensitivity is expected to be about one order of magnitude
  weaker than current LHC limits.  Thus, $Q_{uG}^{k3(3k)}$ is the
  third low-sensitivity operator. We remark, however, that this
  operator plays an important role in global analyses at NLO in QCD
  \cite{durieux}.

\refstepcounter{aux}  
\item[ \theaux] As discussed in section \ref{sec:mmtunn},
  the sensitivity of the MuC top-production processes to $Q_{dW}$ and
  $Q_{\vphi ud}$ is very low, as a result of the their very small
  cross sections. Their contribution to $t\bar q \nu_\mu \bar \nu_\mu$
  production is either zero for $C^{\vphi ud}_{3k}$ and $C^{dW}_{3k}$,
  or nevertheless negligible for $C^{\vphi ud}_{k3}$ and $C^{dW}_{k3}$
  (see eq.~(\ref{eq:tunn})).  Taking into account also that both
  operators get stringent constraints from $B$ measurements, we
  conclude that they can be safely discarded from future studies
  outright.

\refstepcounter{aux}  
\item[  \theaux] Finally, we remark that some operator coefficients are
  associated with couplings that do not involve the top quark.  In
  particular, $C_{1+3}^{\vphi q(-)}$, $C_{1+3}^{\vphi q(+)}$,
  $C_{13}^{\vphi ud}$ and $C_{31}^{uW}$ generate a $Wub$ vertex, and
  $C_{31}^{dW}$ and $C_{13}^{dW}$ generate a $Zbd$ vertex. All these
  operators get strong constraints from $B$ meson measurements.
\end{trivlist}


\paragraph*{Acknowledgments}

We are grateful to Georgina Espinoza Gurriz for her assistance with our
computer hardware.  We acknowlegde support from Sistema Nacional
de Investigadores de Conacyt, M\'exico.  We acknowledge that an arXiv
has previously been published \cite{arxiv}.


\appendix{}

\section{From gauge to mass eigenstates.}
\label{sec:app}

In this work we follow the same definitions of Wilson coefficients
and the Feynman rules as given in ref.~\cite{feynrules} for all operators
except for the operators $Q^{(1)}_{\vphi q}$ and $Q^{(3)}_{\vphi q}$. 
In \cite{feynrules} they use $U_{d_L}$ to redefine the Wilson
coefficients ${C'}^{\vphi q(1)}$  and ${C'}^{\vphi q(3)}$.  In our case
we use $U_{u_L}$ for ${C'}^{\vphi q(-)}$, and $U_{d_L}$ for ${C'}^{\vphi q(+)}$:
\bea
C^{\vphi q(-)} &=& U^{\dagger}_{u_L} {C'}^{\vphi q(-)} U_{u_L} \; ,\;\;\;
C^{\vphi q(+)} = U^{\dagger}_{d_L} {C'}^{\vphi q(+)} U_{d_L}  \; .
\nonumber
\eea
Comparing with the Feynman rule for the $Ztu$
coupling in \cite{feynrules} we have
\bea
C_{13(31)}^{\vphi q(-)} \simeq V_{ud} V_{tb}
({\bf C}_{13(31)}^{\vphi q1}-{\bf C}_{13(31)}^{\vphi q3}) \; ,
\eea
where ${\bf C}_{13(31)}^{\vphi q(1,3)}$ are the Wilson coefficients in
\cite{feynrules} and we have omitted terms proportional to non-diagonal
CKM elements.   As mentioned above, operators $Q_{\vphi u}$ and
$Q^{(\pm)}_{\vphi q}$ are Hermitian.  Therefore, when we rotate to mass
eigenstates the coefficients in $C^{\vphi u}$ and $C^{\vphi q(\pm)}$ are taken
as Hermitian $3\times 3$ matrices.
Then, the matrix elements can be written as (see \cite{durieux}):
\bea
C_{a3}^{\vphi q(\pm)} &=& C_{3a}^{\vphi q(\pm)*} \; \equiv C_{a+3}^{\vphi q(\pm)}
\; , \;\;\;
C_{a3}^{\vphi u} = C_{3a}^{\vphi u*} \; \equiv C_{a+3}^{\vphi u}
\; ,\nonumber 
\eea
for $a=1,2$. The effective Lagrangian is then written as a sum of
operators in mass eigenstate basis along with their coefficients, with
the non-Hermitian operators added their Hermitian conjugate term.

\section{LHC bounds on effective top couplings}
\label{sec:lhc.bounds}

The following limits are obtained from FCNC $Br(t\to Vq)$, $V=g,$
$\gamma$, $Z$, $H$ recent bounds from the LHC \cite{atlas2023}

\bea
&(t\to u\gamma)& |C_{13(31)}^{uB}| \leq 0.037\; ,
\;\;\;\; |C_{13(31)}^{uW}| \leq 0.066
\nonumber \\
&(t\to uZ)& |C_{13(31)}^{uB}| \leq 0.23\; ,
\;\;\;\; |C_{13(31)}^{uW}| \leq 0.13
\nonumber \\
&(t\to uZ)& |C_{1+3}^{\vphi u}| \leq 0.19\; ,
\;\;\;\; |C_{1+3}^{\vphi q(-)}| \leq 0.10
\nonumber \\
&(t\to uH)& |C_{13(31)}^{u\vphi}| \leq 1.2\; ,
\;\;\;\; (t\to ug) |C_{13(31)}^{uG}| \leq 0.074
\nonumber \\
&(t\to c\gamma)& |C_{23(32)}^{uB}| \leq 0.081\; ,
\;\;\;\; |C_{23(32)}^{uW}| \leq 0.15
\nonumber \\
&(t\to cZ)& |C_{23(32)}^{uB}| \leq 0.34\; ,
\;\;\;\; |C_{23(32)}^{uW}| \leq 0.19
\nonumber \\
&(t\to cZ)& |C_{2+3}^{\vphi u}| \leq 0.28\; , 
\;\;\;\; |C_{2+3}^{\vphi q(-)}| \leq 0.14 
\nonumber \\
&(t\to cH)& |C_{23(32)}^{u\vphi}| \leq 1.4\; , 
\;\;\;\; (t\to cg) |C_{23(32)}^{uG}| \leq 0.18
\label{lhclimits}
\eea

\end{document}